\documentclass[twocolumn,twocolappendix]{aastex701}

\usepackage{amsmath} 
\usepackage{chngcntr}
\usepackage{xcolor}
\usepackage{xspace}
\usepackage{afterpage}

\newcommand{\RTnsCRiMHD}{Full-Physics\xspace}

\newcommand{\AZSpaper}{{Martin-Alvarez et al. under review}}
\newcommand{\Msun}{\mathrm{M}_{\odot}}
\newcommand{\SigmaUV}{\Sigma_\text{150W}}
\newcommand{\SigmaUVunits}{{\rm mag\,arcsec^{-2}}}
\newcommand{\magunits}{{\rm mag}}
\newcommand{\Mstar}{\mathrm{M}_{*}}
\newcommand{\Rstellar}{\mathrm{R}_\text{stellar}}
\newcommand{\Rhl}{\mathrm{R}_\text{half-light}}
\newcommand{\Robs}{\mathrm{R}_\text{obs}}
\newcommand{\pc}{\mathrm{pc}}
\newcommand{\kpc}{\mathrm{kpc}}

\begin{document}

\title{Non-Thermal Physics Drives Compact, Self-Regulated Galaxy Morphologies at Cosmic Dawn}
% Comprehensive Physics Bear Compact, Self-Regulated Galaxy Morphologies at Cosmic Dawn
\author[orcid=0009-0009-3348-6688]{Michelle S. Park}
\affiliation{Department of Physics, Stanford University, Stanford, CA 94305, USA}
\affiliation{Kavli Institute for Particle Astrophysics \& Cosmology (KIPAC), Stanford University, Stanford, CA 94305, USA}
\email[show]{michpark@stanford.edu}
\correspondingauthor{Michelle S. Park}

\author[0000-0002-4059-9850]{Sergio Martin-Alvarez}
\affiliation{Kavli Institute for Particle Astrophysics \& Cosmology (KIPAC), Stanford University, Stanford, CA 94305, USA}
\email{martin-alvarez@stanford.edu}

\author[0000-0003-2229-011X]{Risa H. Wechsler}
\affiliation{Department of Physics, Stanford University, Stanford, CA 94305, USA}
\affiliation{Kavli Institute for Particle Astrophysics \& Cosmology (KIPAC), Stanford University, Stanford, CA 94305, USA}
\affiliation{SLAC National Accelerator Laboratory, Menlo Park, CA 94025, USA}
\email{rwechsler@stanford.edu}

\begin{abstract}
James Webb Space Telescope (JWST) has discovered unexpectedly bright, rapidly growing galaxies in the early universe, which were not predicted by most previously existing galaxy formation models.
Using synthetic JWST observations of the \textsc{Azahar} simulation suite, we show that comprehensive non-thermal physics (``Full-Physics'') produces compact, self-regulated galaxies that match observations from $z = 12$ to $z = 3$. 
This model also produces broad surface brightness distributions, where the bright end is dominated by compact sizes and bursty star formation. This compact starburst scenario naturally explains the detection of bright $z > 10$ galaxies in flux-limited surveys.
By contrast, a model with standard hydrodynamics yields systems that are smaller and more concentrated than current data, while a model with calibrated supernova feedback produces unphysically large systems nearly twice the size of those observed.
At lower masses ($\Mstar < 10^{8}\,\Msun$), the \RTnsCRiMHD model predicts sizes that are smaller than can be resolved with JWST, consistent with extrapolations from observations of higher-mass systems. 
Future observations with higher resolution could resolve this population and elucidate the physics driving the formation of the first galaxies.
\end{abstract}

%https://astrothesaurus.org/concept-select/
\keywords{\uat{Galaxy Formation}{595} --- \uat{High-Redshift Galaxies}{734} --- \uat{James Webb Space Telescope}{2291}}

\section{Introduction}
The James Webb Space Telescope (JWST) has found two surprising features of the first galaxies at Cosmic Dawn ($z > 10$): (1) rapid stellar mass assembly within the first 500 million years after the Big Bang \citep{Naidu2022TwoJWST,Labbe2023ABang, Donnan2023TheImaging}, and (2) an unexpectedly high number of UV-bright galaxies \citep{Castellano2022Early915, Harikane2023AEpoch, Finkelstein2023CEERSJWST, Whitler2025TheReionization, Chemerynska2026TheZ9-15}. Previously existing models of galaxy formation anticipated a steep build-up of these galaxies at later times \citep{Cowley2018PredictionsCDM, Yung2019Semi-analytic,Vogelsberger2020High-redshiftFunctions}, informed by pre-JWST observations restricted to the low-redshift regime ($0 < z < 3$) that lacked sensitivity to the fainter galaxies \citep[][]{VanDerWel2012STRUCTURALCANDELS, Lange2015GalaxyMorphology, Shibuya2015MORPHOLOGIESEVOLUTION, Bouwens2015UVFIELDS}. While some pre-JWST observations suggested a build-up of galaxies from $z \sim 10$ to $z \sim 8$ \citep[e.g.,][]{Bradley2012THEFUNCTION, Oesch2012EXPANDED8, Oesch2013PROBINGDATA, Bouwens2014A9.2, Bowler2020A10}, these detections were limited by small sample sizes and observational uncertainties.
JWST has now established bright, extended galaxies at $z > 7$ that challenge the expected timelines for assembly \citep{Suess2022Rest-frameAppeared, Westcott2025EPOCHS12.5, Allen2024GalaxyImaging} and structure formation \citep{Ferreira2022PanicField, Ferreira2023The6.5, Jacobs2023EarlyOptical}. 

%has now established that bright, extended galaxies were already in place at $z>7$, demonstrating that such systems are not exceptional and shifting the inferred timeline of galaxy assembly substantially earlier than expected

The evolution of the size--mass relation has served as a key diagnostic to trace structural assembly and galaxy growth. However, accurately measuring galaxy morphologies at $z > 6$ remains challenging \citep{Kartaltepe2023CEERSJWST,Harikane2025JWSTUniverse}. For these galaxies, measured sizes are near or under the resolution limit of JWST, which biases morphology and size measurements toward PSF properties and impedes the recovery of the intrinsic galaxy profile \citep{Trujillo2001ThePSF,Wang2024TheJWST}. Small and compact galaxies, which are more common at high redshift, are particularly sensitive to these effects \citep{Ono2023Morphologies12} and may overestimate inner radii by up to a factor of five \citep{Wang2024TheJWST}. These biases will also affect morphological parameters characterizing flux distribution, such as concentration, requiring detailed PSF modeling at high redshifts ($z \gtrsim 7$; \citealt{Morishita2024Enhanced14}).

Simulations have served as the primary testbeds to unravel the complex physical processes revealed by observations.
The calibration of star formation and feedback to pre-JWST observations has compensated for limited observational resolution and incomplete feedback physics \citep[e.g.,][]{Agertz2013TOWARDSIMULATIONS, Crain2015TheVariations, Pawlik2017TheResults, Pillepich2018SimulatingModel}. At low redshift ($z \leq 2$), these models have been particularly successful in reproducing observed sizes, masses, and structural diversity \citep[e.g.,][]{Schaye2015TheEnvironments, Dave2016Mufasa:Hydrodynamics, Tremmel2017TheSMBHs, Pillepich2019FirstTime}. However, simulations with supernova (SN) feedback alone have failed to reproduce relations such as the observed stellar mass functions or galaxy halo constraints \citep{Hopkins2014GalaxiesFormation, Smith2019CosmologicalAlone, Katz2023TheSimulations}.  
At high redshift, the limitations of calibration have become more evident with the first light of JWST. Models calibrated to pre-JWST observations often produce galaxies that form too late and fail to reproduce the rapid growth and UV-bright luminosities observed in galaxies at early times \citep{Yung2023AreModels, Kannan2023The8, Robertson2026AcceleratedModels}. To resolve these discrepancies, recalibrating SN parameters is insufficient: introducing more complex calibration and free parameters reduces the predictive efficiency of the simulations and weakens the physical mapping between simulated and observed galaxies.

Magnetohydrodynamics, radiative transfer, and cosmic rays offer a particularly promising solution due to their substantial energy budgets and well-known effects on galaxy evolution. Simulations that incorporate these processes have more closely reproduced observed galaxy properties at both high and low redshift \citep{Pillepich2018SimulatingModel, Chan2019CosmicEmission, Kannan2022IntroducingReionization, Hopkins2025COSMICMEDIUM, Martin-Alvarez2025TheGalaxies}.  
Observations suggest a rough equipartition of magnetic, turbulent, and thermal energies in the interstellar medium (ISM) \citep{Beck2015MagneticStudy, Dacunha2025TheGalaxies}, with strong galaxy magnetizations observed at early times \citep{Bernet2008StrongRedshift, Geach2023Polarized2.6}.
Their presence suppresses gas cooling, fragmentation, and mixing \citep{Tan2013TheAlma, Kauffmann2013LOWFIELDS, Sur2014MixingMedia, Maglione2025}, modulating the efficiency, spatial distribution, and clustering of star formation \citep{Robinson2024RegulatingGalaxy, Belfiori2025OnGalaxies}. As a result, global properties of galaxies such as their stellar masses and disk sizes are affected \citep{Marinacci2015EffectsPopulation, Martin-Alvarez2020HowGalaxies}. 
Radiation provides the highest integrated energy feedback from stars \citep{Leitherer1999Starburst99:Formation}, suppressing the pervasive formation of star-forming clumps \citep{Rosdahl2015GalaxiesGalaxies} and halting star formation after the first massive stars form \citep{Geen2015AStar, Agertz2020}. Despite this local suppression, radiative feedback may increase global star formation due to lower gas ejection through outflows \citep{Smith2021EfficientClustering, Martin-Alvarez2022TheGalaxies, Martin-Alvarez2025TheGalaxies}. 
Hadronic cosmic rays, injected by supernovae remnants, drive gas outflows through smooth pressure gradients \citep{Girichidis2018, Dashyan2020CosmicGalaxies, Ruszkowski2023CosmicClusters, Hopkins2025COSMICMODELS,  Martin-Alvarez2025TheGalaxies}, reducing stellar masses by factors of 2--10 from $z \sim 3$ to $z \sim 1$ \citep{Hopkins2020ButFormation, RodriguezMontero2024TheAnalogues}. Additionally, they modify the gas distribution in the circumgalactic medium and the size of molecular clouds \citep{Butsky2020TheMedium}.

The interplay of these non-thermal feedback processes is complex and remains poorly constrained, yet their combination yields more extended, rotationally supported galaxies \citep{Martin-Alvarez2022TheGalaxies} that naturally reproduce a variety of empirical galaxy relations. This agreement highlights the importance of non-thermal physics and motivates extension of such studies to larger samples spanning a wider range of masses and cosmic epochs. 

Here, we present one of the first studies with the new \textsc{Azahar} cosmological simulations. We compare a comprehensive physical model with non-thermal physics versus two hydrodynamics-only models, specifically one with standard feedback and one with calibrated supernovae, to isolate how feedback regulates galaxy morphologies.
We use mock JWST observations to robustly compare with recent observational results, focusing primarily on the size--mass relation and its evolution across cosmic time from $z = 12$ to $z = 3$. 
% We compare intrinsic and observed morphologies, and detectability of the simulated galaxies. 
% We do not tune model parameters to match observables, ensuring an objective assessment of how feedback physics affects galaxy evolution. 
% We find that comprehensive physics produces structured, compact, and self-regulated galaxies that match observations across cosmic time. These realistic bursty and compact morphologies naturally give rise to high surface brightnesses and explain the observed abundance of bright galaxies at $z > 10$.

The structure of this paper is as follows. In Section~\ref{subsec:methods-azahar-sim}, we briefly describe the \textsc{Azahar} simulations and the three physical models explored in this study. Section~\ref{subsec:methods-galaxy-sample} details our galaxy sample selection. In Section~\ref{subsec:methods-mock-obs}, we describe our methodology for generating mock JWST observations. Section~\ref{subsec:methods-size-measurements} describes how we measure intrinsic and observed galaxy sizes. Section~\ref{sec:results} presents our main findings. We show how different feedback physics models produce visually distinct galaxy morphologies (Section~\ref{subsec:results-dif-morphologies}) and size--mass relations (Section~\ref{subsec:results-size-mass}). We then isolate the effects that produce large early galaxy sizes (Section~\ref{subsec:r-z-results}), and discuss the implications on galaxy detectability at high redshift (Section~\ref{subsec:results-surface-brightness}). We summarize in Section~\ref{sec:conclusion}.

\section{Methods}
\subsection{The \textsc{Azahar} simulations} \label{subsec:methods-azahar-sim}
\textsc{Azahar} is a new suite of high-resolution, zoom-in cosmological simulations designed to study the role of non-thermal processes in shaping galaxy formation and evolution. A detailed description of the numerical methods and physics implementation will be presented in \AZSpaper. Generated using the magnetohydrodynamical adaptive-mesh-refinement code RAMSES \citep{Teyssier2001CosmologicalRAMSES}, \textsc{Azahar} is comprised of multiple models exploring combinations of different physics including hydrodynamics, magnetic fields, radiative transfer, and cosmic rays. The models assume a Planck Cosmology from\citet{PlanckCollaboration2016PlanckFields}: total matter density $\Omega_\text{m} = 0.3065$, baryon density $\Omega_\text{b} = 0.0483$, dark energy $\Omega_{\Lambda} = 0.6935$, and a reduced Hubble constant $h = 0.679$. 
The simulations follow the evolution of several hundred galaxies within a convex-hull zoom-in region spanning approximately 8~comoving Mpc (cMpc), embedded in a larger cosmological box of 25 cMpc per side. The most massive galaxy in the \textsc{Azahar} zoom sub-volume is a spiral galaxy that forms from the merger of two main progenitors at $z \sim 6.5$. This galaxy grows to a halo mass of around $2.5 \times 10^{12}~\Msun$ and a stellar mass of around $6 \times 10^{10}~\Msun$ by $z \sim 1$. 

The high resolution of the \textsc{Azahar} simulations enables the internal structure of galaxies to be resolved well. The target full-cell width spatial resolution is $\Delta x = 23.8~\text{pc}$. The high-resolution dark matter particles have masses of $m_{\text{DM}} \approx 4.5 \times 10^5~\Msun$, and the stellar particles have masses $m_{\text{*}} \approx 4 \times 10^4~\Msun$. Additionally, the models include a Jeans length refinement criterion in combination with Lagrangian mass refinement \citep{Martin-Alvarez2022TowardsGalaxies}, ensuring better resolution of multi-scale ISM turbulence and magnetic vector fields.

In this study, we compare three models, each with a different physical prescription: (1) HD with standard hydrodynamics and feedback, (2) HD-Boost with enhanced supernovae feedback, and (3) ``\RTnsCRiMHD'' with radiative transfer, cosmic rays, and magnetism. All models include radiative cooling above and below $10^4~\text{K}$ \citep{Rosen1995GlobalGalaxies, Ferland1998CLOUDYSpectra}, a magneto-thermoturbulent prescription for star formation \citep{Federrath2012THEOBSERVATIONS, Kimm2017Feedback-regulatedReionization, Martin-Alvarez2020HowGalaxies}, and mechanical supernova feedback \citep{Kimm2014ESCAPESTARS}. Each supernova explodes with an energy of $E_\text{SN} = 10^{51}$~erg generated by a progenitor of $\rm{M}_\text{SN} = 10~\Msun$. For the HD-Boost model, we increase supernova energy injection by a factor of four, calibrated to reproduce the galaxy stellar mass function at $z = 3$. The HD and \RTnsCRiMHD models are not calibrated.

The \RTnsCRiMHD model additionally includes the following non-thermal physics: 
\begin{enumerate}
    \item Lyman Continuum on-the-fly radiative transfer from stellar sources using the implementation by \citet{Rosdahl2015ARamses-rt} following its configuration in the \textsc{Sphinx} simulations \citep{Rosdahl2018TheReionization}.
    \item Cosmic rays, following the implementation by \citet{Dubois2016TheFeedback}, and sourced by SN feedback that injects 10\% of the total energy of each event as cosmic rays \citep{Morlino2012StrongRemnant}. We do not account for cosmic ray streaming, and assume a constant anisotropic diffusion coefficient of $3 \times 10^{28}~\text{cm}^2 \text{s}^{-1}$ \citep{Salem2016RoleMedium, Pakmor2016Semi-implicitMesh}. 
    \item Magnetic fields, through the constrained transport implementation by \citet{Teyssier2007AMHD}, and astrophysically seeded through magnetized SN feedback, which injects 1\% of the SN energy as magnetic. 
\end{enumerate}
% Through these physics, \textsc{Azahar} offers detailed treatment of the ISM and feedback physics, enabling the study of how its different models influence the growth of galaxies, their sizes, morphologies, and appearance. 

\subsection{Simulated galaxy sample} \label{subsec:methods-galaxy-sample}
We analyze the evolution and properties of galaxies from $z = 12$ to $z = 3$, selecting systems with stellar masses greater than $4 \times 10^6~\Msun$. We focus primarily on galaxies with masses $10^7$--$10^{10}$ M$_{\odot}$, which are well-resolved and statistically well-sampled. 

To track and characterize galaxies in our simulations, we employ new galaxy trackers described in \AZSpaper. This algorithm first employs a dark matter halo catalog generated with \textsc{Halomaker} \citep{Tweed2009BuildingSimulations} to seed galaxy trackers in the centers of halos and subhalos with masses greater than $\text{M}_\text{halo} > 10^8~\Msun$ (v4.7). In our galaxy tracking framework, each subhalo is either required to have an associated galaxy or to seed a new galaxy tracker object. We track galaxies throughout cosmic time by following their innermost stellar particles across snapshots, with a minimum threshold of 50 particles.

% fp: z = 3: 986, z = 4: 839, z = 5: 669, z = 6: 497, z = 7: 391, z = 8: 161, z = 9: 479, z = 10: 224, z = 11: 89, z = 12: 89
% HD: z = 3: 992, z = 4: 791, z = 5: 583, z = 6: 425, z = 7: 264, z = 8: 162, z = 9: 487, z = 10: 206, z = 11: 91, z = 12: 91

Our sample yields $\sim$ 100 galaxies at $z \sim 12$, gradually increasing through intermediate redshifts to several hundred galaxies at $z \sim 7–5$, and reaching $\sim$ 1000 galaxies by $z = 3$ in each of the three models (HD, HD-Boost, \RTnsCRiMHD). 

\subsection{Mock Observations} \label{subsec:methods-mock-obs}
Most of our measurements rely on various 2D projections. We characterize the intrinsic distribution of stellar mass by computing maps of integrated stellar surface density. 
For direct comparison with observational data, we generate mock observations in three JWST/NIRCam filters: F150W, F277W, and F444W. We compute these by integrating the emission of each star, applying wavelength-dependent filter transmissivity. This returns the emission for each filter as it would be observed by JWST. We treat each stellar particle as a single stellar population, with its emission determined by  \citet{Bruzual2003Stellar2003} models according to mass, metallicity, and age.   
We account for dust obscuration along the line of sight by propagating stellar emission through a 3D absorption cube. We apply an exponential extinction law with $R_v \sim 3.1$ \citep{Weingartner2001DustCloud}. We estimate per-cell dust content through an ionization-modulated dust-to-metal ratio following $\eta_\text{D/M}\,(x_\text{HI} + f_\text{ion}x_\text{HII})$, where $x_\text{HI}$ and $x_\text{HII}$ are the neutral and ionized hydrogen mass fractions respectively \citep{Laursen2009LyGalaxies}. We set $\eta_\text{D/M} = 0.4$ and $f_\text{ion} = 0.01$ (see Appendix~A in \citealt{Martin-Alvarez2024} for additional details).

To reproduce instrumental effects, we convolve each synthetic observation with the JWST/NIRCam point-spread function (PSF) for the corresponding filter, and employ the full width at half maximum (FWHM) provided by the STScI NIRCam performance documentation\footnote{\url{https://jwst-docs.stsci.edu/jwst-near-infrared-camera/nircam-instrumentation/nircam-filters}}. For each redshift, we compute the angular diameter distance according to the cosmological parameters of the simulation. We then use this distance to convert the angular PSF FWHM into a physical scale, and compute the effective PSF width at that redshift. We convert this FWHM to the Gaussian width and convolve each synthetic observation with a 2D Gaussian kernel of that width. We discuss the impact of a Gaussian PSF assumption on our size measurements in Appendix~\ref{ap:psf-shape}.

\subsection{Size Measurements} \label{subsec:methods-size-measurements}
To compute galaxy size more consistently with observational studies, we measure circularized, projected 2D sizes directly from our synthetic images. For each galaxy, we first identify the center of the radial light profile by locating the brightest pixel near the tracker position. 
To avoid size biasing, if nearby sources are present in an image, we truncate the radial light profile at half the distance to the nearest neighbor, according to its tracker position.
At shorter separations, light profiles may overlap significantly. Therefore, when two galaxies have a projected physical distance lower than 5 kpc, we fit a two-component Gaussian model to de-blend the light distribution. 

We define three radii measurements: (1) Stellar radius, $\Rstellar$, enclosing 50\% of the total stellar mass in the stellar surface density maps; (2) Half-light radius, $\Rhl$, enclosing 50\% of the per-filter total flux in the mock-observation maps, before applying the JWST/NIRCam PSF; and (3) Observed radius, $\Robs$, the half-light radius computed from the same maps, after applying PSF convolution. For simplicity, we do not include pixel response/drizzling, detector and background noise, or source detection and segmentation in this work. 

We focus our analysis on F444W size except for Section~\ref{subsec:results-surface-brightness}. We compute mass and luminosity measurements from these images as the integrated quantities within $2 \times \Rstellar$ and $2 \times \Rhl$, respectively. 

To infer sizes close to and below instrumental resolution, observational pipelines often rely on forward modeling, assuming a parametric light profile model. In this regime, the recovered sizes are particularly sensitive to effects such as PSF characterization, signal-to-noise ratios, the accuracy of galaxy centering, and irregularities in the light distribution.
As a result, an accurate observational reconstruction is expected to yield estimated sizes lying between $\Rhl$ and $\Robs$.

\section{Results} \label{sec:results}
\subsection{Different galaxy morphologies under different feedback physics models} \label{subsec:results-dif-morphologies}

We first review morphological variations across the simulations, which feature a wide variety of appearances and shapes. As a result, summarizing galaxy morphologies into a subset of parameters does not reflect their complexity. We highlight the most important differences by qualitatively comparing galaxy morphologies in our JWST RGB synthetic observations. 

%%%%%%%%%%% Azahar-a evolution %%%%%%%%%%%
\begin{figure*}[ht!]
    \centering
    \includegraphics[width=1\textwidth]{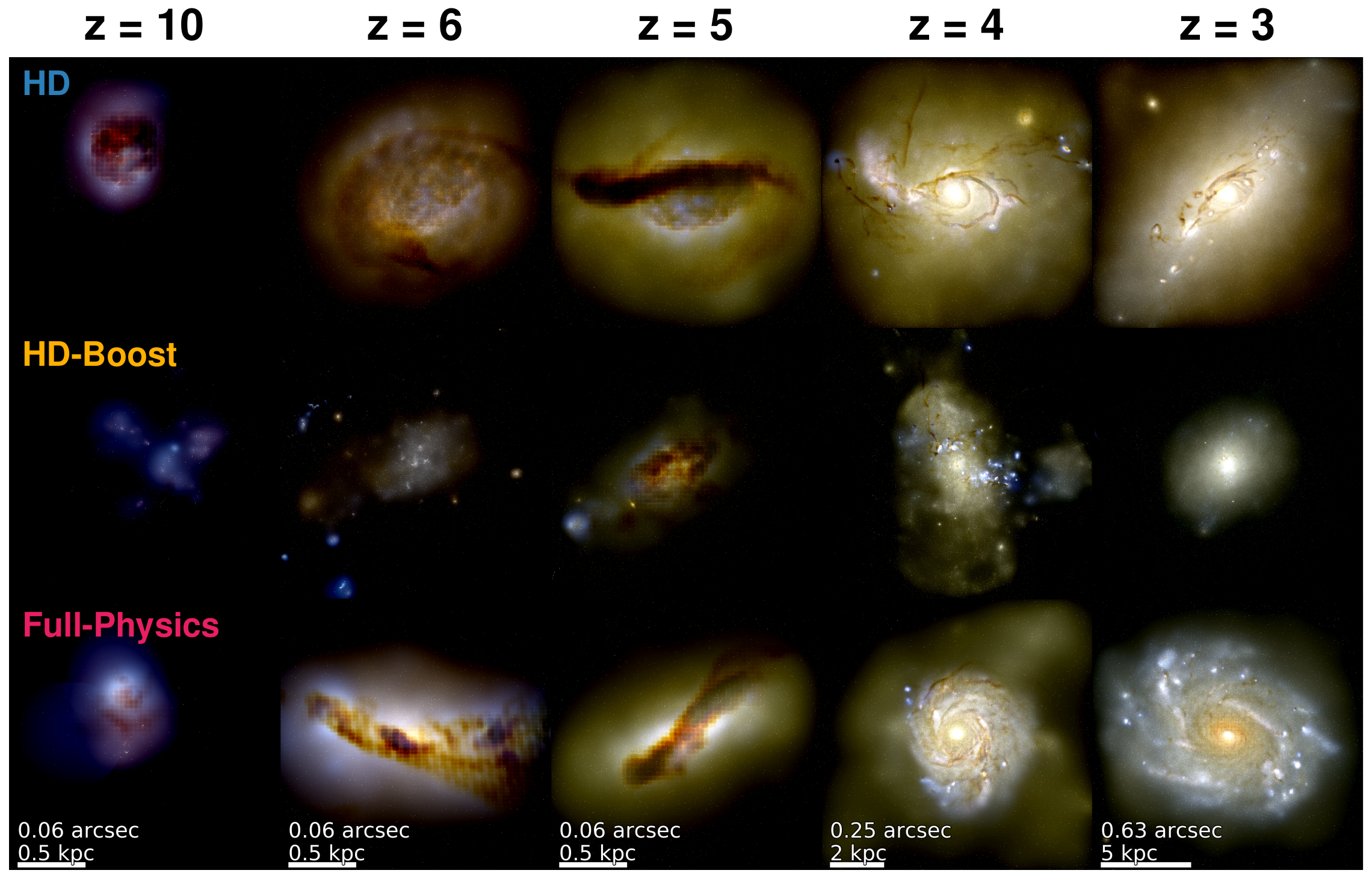}
    \caption{Synthetic observations of Azahar-a, the largest galaxy in our sample, in the JWST filters F150W (blue), F277W (green), and F444W (red). Columns show $z = 10$, $6$, $5$, $4$, to $3$ with panel aperture sizes of $2$, $2$, $2$, $10$, and $15$ physical kpc, respectively. Scale bars indicate physical and angular sizes. Each row from top to bottom corresponds to the following models: standard hydrodynamics (HD), enhanced SN feedback hydrodynamics (HD-Boost), and our \RTnsCRiMHD model with radiative transfer, cosmic rays, and magnetic fields. Different feedback models produce distinct morphological and structural evolution. By $z = 3$, the \RTnsCRiMHD model develops a well-defined disk with spiral arms and dust lanes, while HD-Boost remains diffuse and HD shows an over-concentrated bulge.}
    \label{fig:azahar-a-evol}
\end{figure*}

Figure~\ref{fig:azahar-a-evol} displays the evolution of Azahar-a, the most massive galaxy in our sample, from shortly after the emergence of the first galaxies at $z = 10$ to the beginning of ``Cosmic Noon'' at $z = 3$. We generate synthetic observations of its evolution in the JWST F150W (blue), F277W (green), and F444W (red) filters, shown here without instrumental effects. From left to right, the columns correspond to redshifts $z = 10$, $6$, $5$, $4$, and $3$, selected to capture key stages of the galaxy's growth. As the galaxy size increases, we vary the panel aperture sizes from the leftmost to the rightmost column. 
Rows show the three models studied: standard hydrodynamics (HD), boosted supernova feedback (HD-Boost), and comprehensive physics with radiative transfer, cosmic rays, and magnetic fields (\RTnsCRiMHD).

The morphological evolution of the Azahar-a galaxy is notably different between the different feedback models. At early times ($z \sim 10–6$), all three models have galaxies with strong F150W emission, driven by bright UV emission from high specific star formation activity. The higher dust content in the HD model gives its galaxy a notably redder appearance. The HD-Boost model has the faintest galaxy, due to the effective growth suppression of its stronger feedback. From $z \sim 7$ to $z \sim 6$, Azahar-a experiences a major merger event. In the HD model, merger dynamics drive gas to swirl and accumulate around the core of the galaxy. Through efficient gas retention, the HD galaxy rapidly develops a disk structure in the aftermath of this merger. In contrast, the HD-Boost model exhibits a patchier gas distribution during mergers, and fails to accrete gas efficiently into the main progenitor. As a result, the galaxy remains small and irregular, with multiple small companions that have failed to coalesce by $z = 6$. In the \RTnsCRiMHD model, feedback outflows are less violent and more isotropic around the main galaxy. Once the merging satellites incorporate into the main galaxy, the accreted material builds up a disk with strong bi-conical outflows. This large-scale feedback also clears obscuring dust by $z = 6$ to reveal star formation.

During the period from $z=5$ to $z=4$, the galaxy experiences only minor mergers. This allows the system to assemble into a rotationally supported disk in the HD and \RTnsCRiMHD models. The HD model shows the gradual build-up of a disk, evidenced by the dark dust ring at $z = 6$ and edge-on disk obscuration at $z = 5$. At $z = 4$, the disk exhibits star-forming spiral features, with multiple smaller companions. The \RTnsCRiMHD model develops a well-defined disk with distinct spiral arms and dust lanes by $z = 4$. At $z \lesssim 5$, both HD and \RTnsCRiMHD galaxies are embedded in extended and diffuse stellar halos, resulting from the effective disruption of accreted satellites. In contrast, the HD-Boost model shows an irregular and fragmented system, with no signs of a disk or an extended stellar halo.

By $z = 3$, in the HD model, Azahar-a has developed a disk that is fully rotationally supported. The galaxy in this model has an extended disk with bright star-forming clumps and a less-defined spiral structure. The galaxy is surrounded by several small satellites and has built a prominent concentrated central bulge, characteristic of galaxy formation models with unregulated star formation \citep{Brooks2015BulgeSimulations}. 
By contrast, the galaxy in the HD-Boost model remains smaller in size and does not develop a disk. 

In the \RTnsCRiMHD model, Azahar-a has developed a well-defined disk by $z = 3$, with spiral arms permeated by dust lanes, and star-forming regions bright in near-UV emission (blue; F150W). In this model, minor mergers efficiently accrete onto the disk, which undergoes little disruption. The presence of a well-developed disk collimates the escape of outflows perpendicularly to its plane. 

The appearance of Azahar-a at $z = 3$ in the HD and \RTnsCRiMHD models resembles some observed galaxies. The HD model appears similar to the local galaxy NGC 4753 \citep{Dewangan1999Dust4753}, with absorption features, diffuse disks, and a bright red central bulge. The \RTnsCRiMHD model resembles the local galaxy NGC 4414 \citep{DeBlok2014HALOGASPressure}, with tight spiral arms traced by dust lanes and dotted with numerous bright, blue star-forming clumps. The galaxy in the \RTnsCRiMHD model is also reminiscent of disk galaxy observations from \citet{Elmegreen2009CLUMPYIRREGULARS} and \citet{Sattari2023FractionEnvironment}. 

Galaxy sizes can be estimated through multiple measurements, such as the standard half-light (or half-mass) size. This provides a good compromise between simplicity and adaptability. However, half-light radii are highly sensitive to galaxy concentration. Driven by its pronounced central bulge, the HD model is more concentrated than \RTnsCRiMHD\footnote{Following \citet{Conselice2003TheHistories} and \citet{Ferrari2015MorfometrykaGalaxies}, we define concentration as $C = 5 \times \log_{10}({\text{R}_{90}/\text{R}_{50}})$. We compute $\text{R}_{50}$ and $\text{R}_{90}$ as the radius enclosing 50\% and 90\% respectively of the total flux in the radial profiles from the mock observation maps prior to PSF convolution.}. Among galaxies that subtend comparable extensions in the sky, galaxies with more centrally concentrated profiles will yield smaller half-light radii than those without. 
Therefore, visual inspection may be misleading, and size differences between the models should be assessed through quantitative comparisons. 

%%%%%%%%%%% Repgals IN z = 3 %%%%%%%%%%%
\begin{figure*}[ht!]
    \centering
    \includegraphics[width=\textwidth]{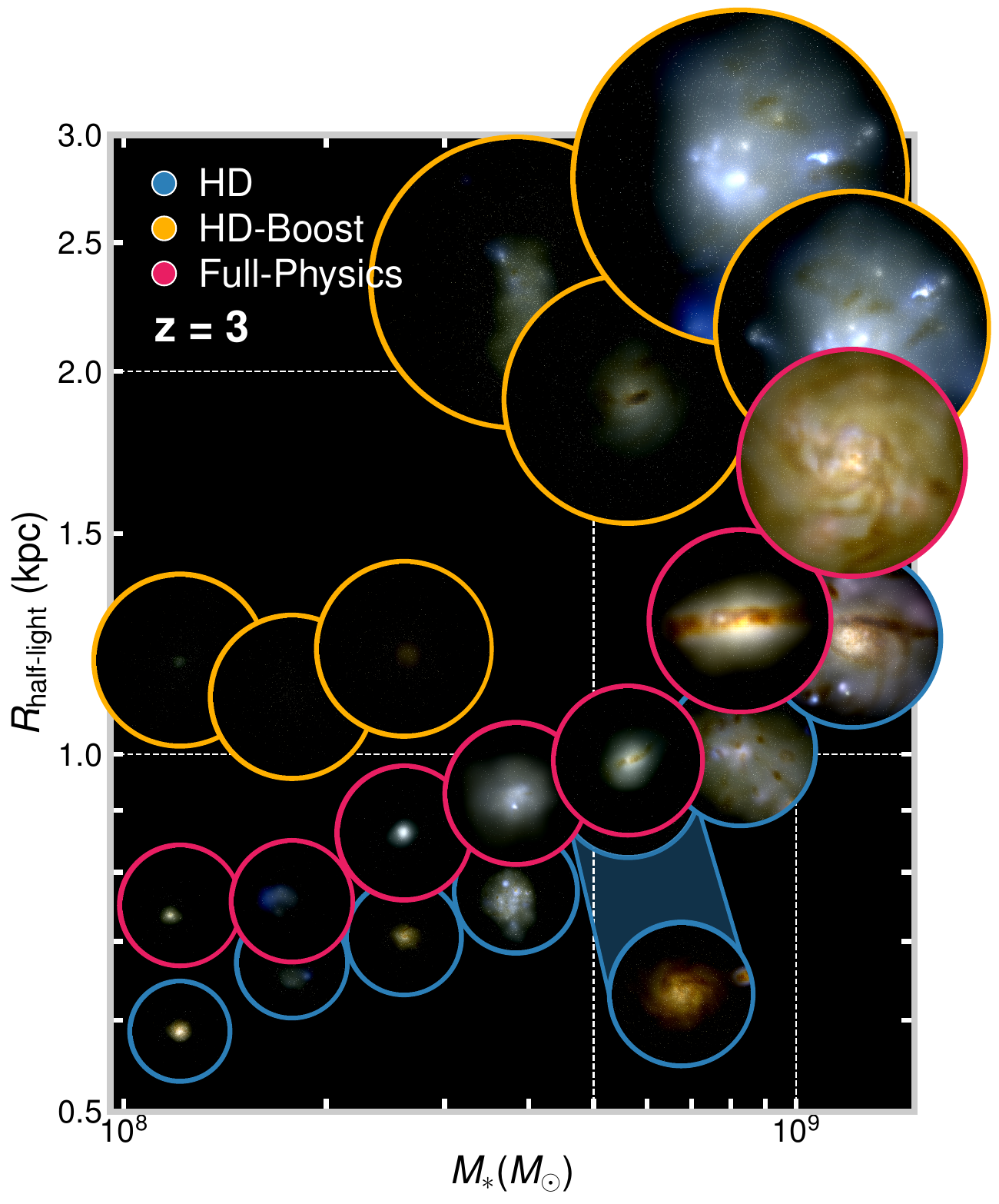}
    \caption{Size--mass relation for the three Azahar models at $z \sim 3$, with representative mock JWST galaxy images. Each median point is placed and proportionally scaled with the galaxy’s $\Rhl$. Synthetic observations use JWST filters F150W (blue), F277W (green), F444W (red). Each of the models are represented by points with differently colored marker edges: standard hydrodynamics (HD; blue), boosted SN feedback (HD-Boost; yellow), and our comprehensive physics model (\RTnsCRiMHD; red). Medians are computed for seven stellar mass bins from $10^8~\Msun$ to $1.5 \times 10^9~\Msun$, representing the high-mass end of our sample. We include dashed white lines for visual reference. HD-Boost galaxies are notably larger and diffuse, HD galaxies remain centrally concentrated, and \RTnsCRiMHD galaxies are larger than HD but within a regulated size regime.}
    \label{fig:repgals-dots}
\end{figure*}

To connect the diverse morphologies and sizes produced by the Azahar simulation with the size--mass relation, we show the relation at $z = 3$ in Figure~\ref{fig:repgals-dots}. Specifically, we show the half-light sizes for the F444W filter for the three models, prior to PSF convolution. The points represent median sizes for seven stellar mass bins spanning $10^8~\Msun$ to $1.5 \times 10^9~\Msun$, for which morphological differences across models are most pronounced. Each data point includes a mock JWST observation (F150W, F277W, F444W) of an individual galaxy with the image scaled proportionally to the galaxy's observed half-light size. 

In the low-mass bins ($10^{8}~\Msun$ to $3 \times  10^{8}~\Msun$), HD galaxies are small and centrally concentrated, often with bright older populations in F444W. \RTnsCRiMHD galaxies are slightly larger than their HD counterparts. By contrast, HD-Boost galaxies are significantly larger, and appear diffuse without distinct stellar cores.

In the four highest mass bins ($3\times 10^{8}~\Msun$ to $1.5 \times 10^{9}~\Msun$), morphological differences become more pronounced. HD galaxies have both blue star-forming clumps and dark dust clumps distributed throughout their disks. \RTnsCRiMHD galaxies are moderately larger than HD, occupying intermediate physical sizes between HD and HD-Boost. HD-Boost galaxies have larger sizes and display highly extended and diffuse morphologies with lower surface brightnesses. 

%%%%%%%%%%% Size-mass across time %%%%%%%%%%%
\subsection{Redshift evolution of the size--mass relation across different physical models} \label{subsec:results-size-mass}
We now explore the size--mass relation and its evolution quantitatively to assess how accounting for additional physical processes affects galaxy sizes and how observational biases alter interpretations of galaxy growth. 

\begin{figure*}[ht!]
    \centering
    \includegraphics[width=0.495\textwidth]{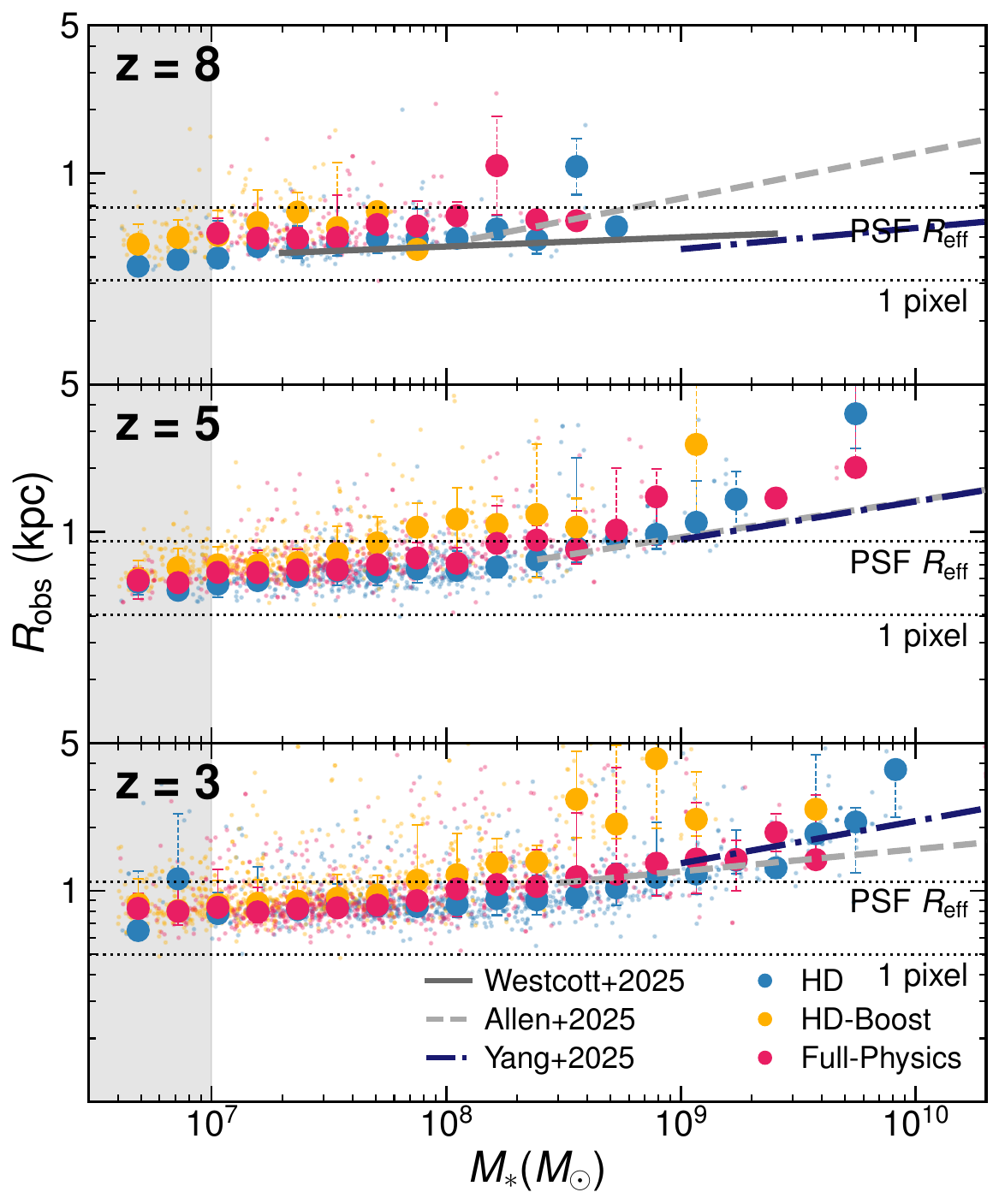}
    \includegraphics[width=0.495\textwidth]{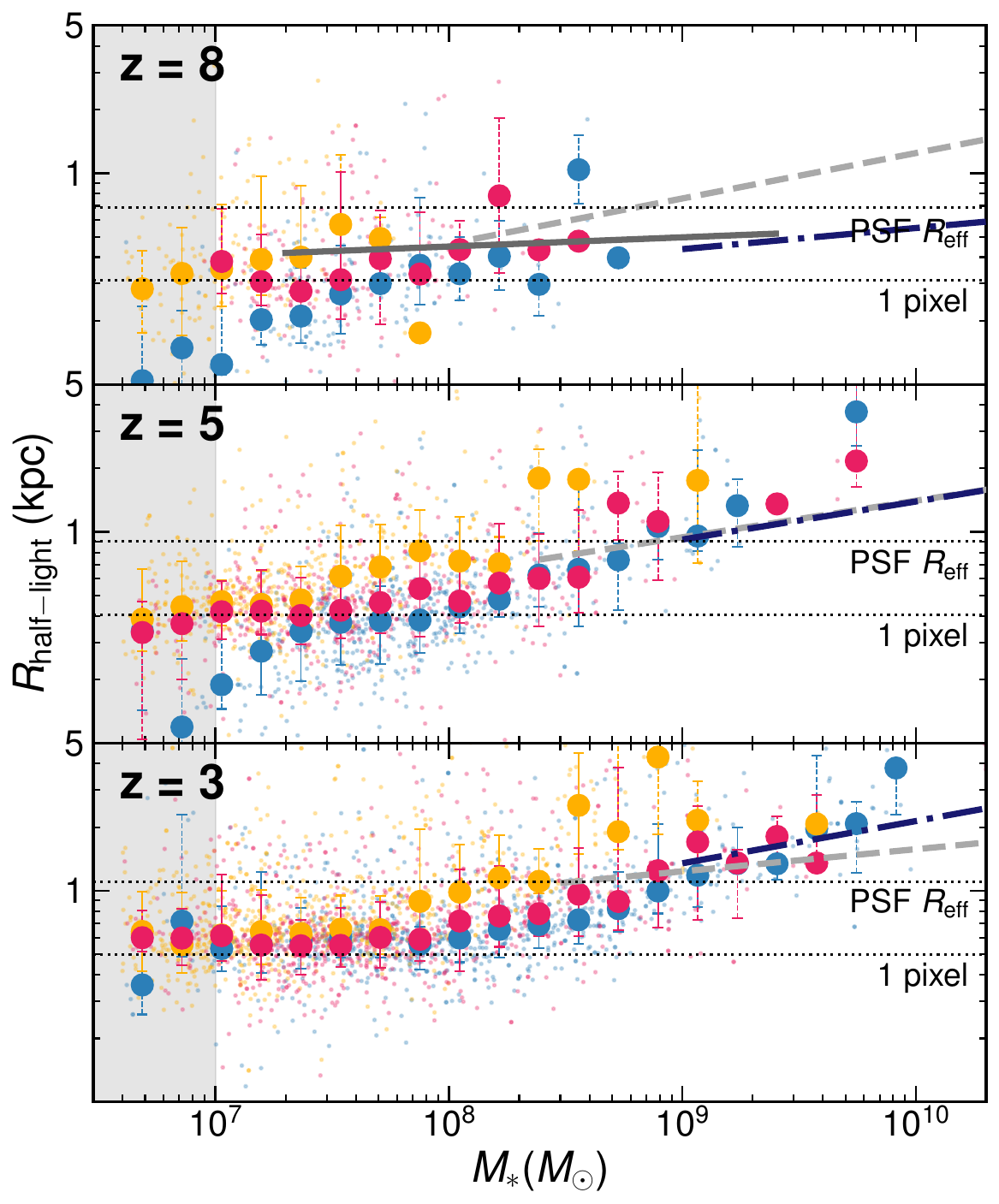}
    \caption{Size--mass relations for PSF-convolved half-light radii $\Robs$ (left column) and half-light radii before PSF convolution $\Rhl$ (right column). From top to bottom, we show redshifts $z = 8$, $5$, and $3$ respectively. Colors indicate models: HD (blue), HD-Boost (yellow), and \RTnsCRiMHD (red). Points show medians and error bars for the 16--84th percentile range. Dashed error bars denote bins with fewer than 20 galaxies. Horizontal black dotted lines mark the JWST/NIRCam F444W PSF physical size and 1 pixel size at the given redshift. Observations are shown for comparison (see text for details).
    %
    %EPOCHS (dark-gray solid; \citet{Westcott2025EPOCHS12.5}), CEERS/PRIMER-UDS/PRIMER-COSMOS (light-gray dashed; \citet{Allen2024GalaxyImaging}), and COSMOS-WEB (navy dash-dotted; \citet{Yang2025COSMOS-Web:2ltzlt10}). 
    %
    The gray shaded region indicates our low resolution limit (galaxies with $\lesssim 300$~stellar particles).  
    At higher masses ($\Mstar > 10^8~\Msun$), the boosted supernova feedback model (HD-Boost) produces extended galaxy sizes, while the \RTnsCRiMHD model regulates sizes between HD and HD-Boost. 
    At lower masses ($\Mstar < 10^8~\Msun$), our models predict that $\Robs$ galaxy sizes are unresolved. The \RTnsCRiMHD model best agrees with the extrapolated $\Rhl$ sizes from higher-mass, well-resolved systems across all redshifts.}
    \label{fig:size-mass-evol}
\end{figure*}

Figure~\ref{fig:size-mass-evol} shows the size--mass relations across our three models, with panels for $z = 8$, $5$, and $3$ (top to bottom). The left column shows the PSF-convolved half-light radius $\Robs$ as a function of stellar mass and the right column shows the half-light radius prior to PSF convolution $\Rhl$. The solid points show the median sizes in 20 stellar mass bins across our sample. Error bars mark the 16th and 84th percentiles, which are shown as dashed error bars when a bin contains fewer than 20 galaxies. The dotted lines indicate the physical size of the JWST/NIRCam F444W PSF FWHM ($0.140\arcsec$) and 1 pixel at the given redshift ($0.063\arcsec/\mathrm{pixel}$).
The gray shaded region indicates our low resolution limit for galaxies with $\lesssim 300$~stellar particles. The most massive galaxies exhibit larger scatter due to small sample sizes. We underlay as faint points the measurements for individual galaxies, color-coded by the corresponding simulation. For comparison, thick lines show the results from observational studies that compute sizes from F444W imaging: the high-redshift ($7.5 <z < 8.5$) EPOCHS survey data \citep{Westcott2025EPOCHS12.5} (top panel); the COSMOS-Web survey \citep{Yang2025COSMOS-Web:2ltzlt10} at $z = 7.5$ (top panel), $z = 4.5$ (central panel), $z = 2.5$ (bottom panel); and the results from \citet{Allen2024GalaxyImaging} for the CEERS, PRIMER-UDS, PRIMER-COSMOS surveys at $6 \leq z < 9$ (top panel), $5 \leq z < 6$ (central panel), and $3 \leq z < 4$ (bottom panel). \citet{Ormerod2024EPOCHSObservations} and \citet{Miller2025JWSTUniverse} also use F444W-based sizes at $z \sim 8$, which corresponds to rest-frame optical emission. Filter choice affects the inferred sizes across redshifts, and studies using alternative configurations \citep[e.g.,][]{VanDerWel20143D-HST+CANDELS:3, Morishita2024Enhanced14, Varadaraj2024ThePRIMER, Ward2024EvolutionCEERS, Genin2025DAWNType, Carreira2026JWSTGOODS-S} can yield informative but correspondingly different size--mass relations. We restrict our comparisons to F444W-based measurements across cosmic time for consistency. We provide double power-law fitting results in Appendix~\ref{ap:size-mass-fits}. 

From $z = 8$ to $z = 3$, galaxies exhibit a systematic increase in size, as shown by the rising median sizes at fixed stellar mass. Each model drives galaxy growth differently, with size differences across models being most pronounced at larger masses ($\Mstar > 10^{8}~\Msun$) and earlier in cosmic time. The galaxies in the HD and \RTnsCRiMHD models grow rapidly in mass at early times ($z > 5$). Conversely, HD-Boost consistently yields the largest galaxies for a given mass, and across all redshifts. Due to its enhanced supernova feedback, this model enhances galaxy sizes, especially toward the high-mass end. 

To evaluate whether the models reproduce realistic galaxy sizes, we compare them first in the left column with observational data. At $z = 8$, the HD model appears to match closely with EPOCHS results. This agreement is driven by the highly concentrated sizes of HD galaxies. The intrinsic half-light sizes of these galaxies are well below the physical size of the PSF, and are thus PSF-dominated. 
This unresolved effect is illustrated by comparing sizes across columns: PSF-convolved sizes are systematically larger, particularly at the low-mass end where the size--mass relation flattens. As discussed in Section~\ref{subsec:methods-size-measurements}, for galaxies with intrinsic sizes between PSF~$\text{R}_\text{eff}$ and 1 pixel size, a non-biased reconstruction model would recover values between $\Robs$ (left column) and $\Rhl$ (right column). Below 1 pixel size, the result is primarily determined by the reconstruction model. 

When comparing our results with the observations by \citet{Allen2024GalaxyImaging}, we find good agreement both across the shared mass range (i.e., $\sim 2 \times 10^{8}\,\Msun$) and when extrapolating both our simulated and their observational results to lower masses. These observations target more massive and extended galaxies that subtend larger angular sizes, and are therefore less susceptible to PSF effects. At $z = 8$, we find that the \citet{Allen2024GalaxyImaging} relation deviates from EPOCHS, predicting a steep upturn in galaxy sizes at higher masses ($\Mstar > 10^8~\Msun$).  

We now examine how these trends evolve at lower redshift. In the left column of $z = 5$, the HD model matches observations by \citet{Allen2024GalaxyImaging} most closely, again due to PSF effects setting an approximate size floor. Without PSF convolution (right column), the \citet{Allen2024GalaxyImaging} relation aligns better with the \RTnsCRiMHD model, even when extrapolated to lower masses ($\Mstar < 10^8~\Msun$). The HD model follows as the next closest match. At this redshift, COSMOS-Web results \citep{Yang2025COSMOS-Web:2ltzlt10} closely follow the \citet{Allen2024GalaxyImaging} relation.

At lower redshifts ($z = 3$), the differences in the observational size--mass relations between the models become less pronounced as galaxies grow to become better resolved. Yet, the HD-Boost model shows a sharp size upturn at high masses, likely due to adiabatic expansion under repeated feedback cycling \citep{El-Badry2016BREATHINGGALAXIES}. Conversely, at the high-mass end, \RTnsCRiMHD aligns most closely with observations, through its regulated galaxy sizes between the more diffuse systems of HD-Boost and the compact galaxies in HD. At this redshift, both the \citet{Allen2024GalaxyImaging} and \citet{Yang2025COSMOS-Web:2ltzlt10} results align most closely with this model, both with and without instrumental effects.

Altogether, these results point toward two distinct scenarios for low-mass galaxy sizes at $z \gtrsim 8$: (1) galaxies are small and compact ($\Rhl < 0.5\,\kpc$) with their star formation concentrated into a small number of massive clumps; or (2) galaxies are extended and diffuse ($\Rhl \gtrsim 0.6\,\kpc$), with multiple and fragmented star-forming regions. 
The compact scenario is favored by both our \RTnsCRiMHD and HD models, and is consistent with the size--mass scaling observed by \citet{Allen2024GalaxyImaging}. Conversely, the extended scenario is the result of enhanced supernova feedback in HD-Boost.

%%%%%%%%%%% Redshift-r 2 panels (3 models, 3 radii) %%%%%%%%%%%
\subsection{Comparing size evolution across cosmic time, shaped by observational limitations} \label{subsec:r-z-results}
To better characterize how galaxy sizes evolve across cosmic time, we consider the global size evolution across redshift, before separating its evolution across galaxy selections such as different stellar masses.

\begin{figure*}[ht!]
    \centering
    \includegraphics[width=\textwidth]{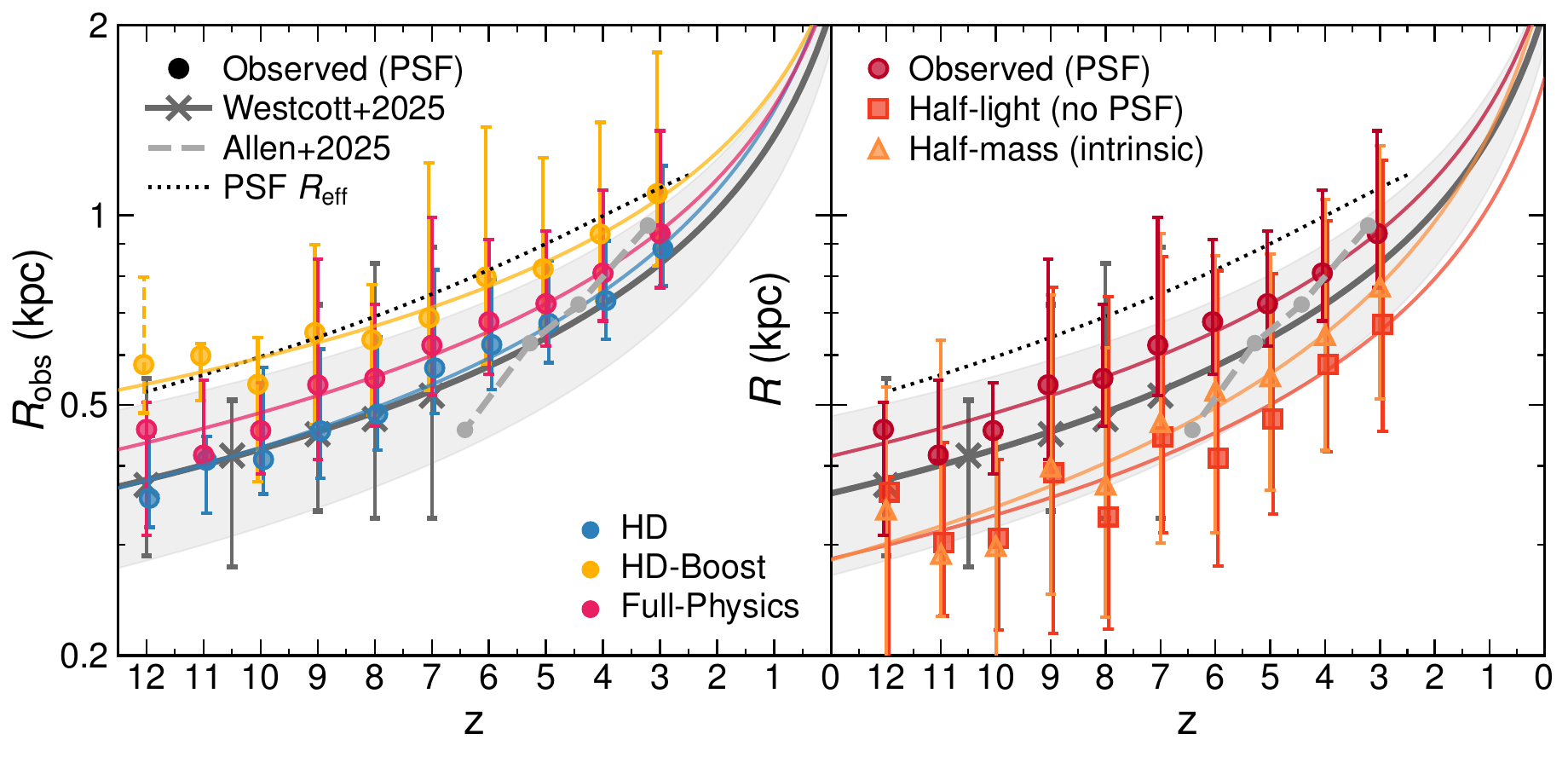}
    \caption{(Left panel) PSF-convolved galaxy half-light radius evolution ($\Robs$) as a function of redshift for our three models: HD (blue), HD-Boost (yellow), and \RTnsCRiMHD (red). Median galaxy sizes are weighted by the EPOCHS mass distribution. Error bars indicate the 16th and 84th percentiles. Colored solid lines show best-fit power-laws, provided in Appendix~\ref{ap:z-r-fits}. The black dotted line shows the physical full width at half maximum size of the JWST/NIRCam F444W PSF. For comparison, we show the EPOCHS results \citep{Westcott2025EPOCHS12.5} as the dark-gray solid line with shaded region, and the EPOCHS-weighted data (see text for details) from \citet{Allen2024GalaxyImaging} as the light-gray dashed line. 
    The HD-Boost model produces unrealistically large observed sizes, whereas the HD model aligns well with the data by EPOCHS. However, the highly concentrated galaxy morphologies in this model are most susceptible to PSF effects. 
    The \RTnsCRiMHD model regulates sizes slightly above EPOCHS, and show better agreement with the extrapolation from the more massive galaxies. 
    (Right panel) Comparison of the \RTnsCRiMHD size evolution for three size definitions: intrinsic half-mass radius $\Rstellar$ (triangles), half-light radius prior to PSF convolution $\Rhl$ (squares), and the PSF-convolved half-light radius $\Robs$ (circles). Galaxy sizes increase systematically under PSF effects, leading to galaxy sizes slightly above EPOCHS. Data from massive galaxies align most closely with the intrinsic galaxy sizes at high redshift.}
    \label{fig:z-r-with-literature}
\end{figure*}

The evolution of median galaxy half-light radii over cosmic time for our models, spanning $z=12$ to $z=3$, is shown in the left panel of Figure~\ref{fig:z-r-with-literature}.  Our primary comparison is with the observations from the EPOCHS survey \citep{Westcott2025EPOCHS12.5}. We fit a Gaussian probability density function (PDF) to the mass distribution at each redshift in the EPOCHS sample. For $z < 7$, we continue to adopt their mass distribution at $z = 7$. We use these PDFs to mass-weight our model size distributions, enabling a more robust comparison. We further discuss the effects of mass-weighting to the EPOCHS sample across redshift in Appendix~\ref{ap:epochs-matching}.
We fit the radius--redshift relation to a standard power law. Our fits are further detailed in Appendix~\ref{ap:z-r-fits}. 
In addition, we show the results from CEERS, PRIMER-UDS, and PRIMER-COSMOS \citep{Allen2024GalaxyImaging} as the dashed light-gray line and circular points. This $R\,(z)$ relation is reconstructed from the F444W size--mass relations at the median redshift of each redshift bin between $3 < z < 9$ and weighted by the EPOCHS mass distribution. Finally, we show the physical full width at half maximum size of the JWST/NIRCam F444W PSF. 

Across all redshifts, the boosted supernova feedback model (HD-Boost) consistently yields the largest galaxies and the greatest size scatter. While the HD-Boost size distributions are not inconsistent with the EPOCHS data, the median HD-Boost galaxy sizes are well above the upper $1\sigma$ error bounds of the observations. As described before, the \RTnsCRiMHD model produces galaxies larger than those in HD, but regulated below the unphysically large sizes of HD-Boost. 
The HD model has the least size scatter, and is the closest match to the EPOCHS relation. However, it is subject to the point-spread function limitations described for highly concentrated galaxies.

To further explore these instrumental effects, we show in the right panel of Figure~\ref{fig:z-r-with-literature} the redshift evolution of different galaxy size definitions, only for the \RTnsCRiMHD model. These sizes are the intrinsic half-mass radius $\Rstellar$, half-light radius prior to PSF convolution $\Rhl$, and the PSF-convolved half-light radius $\Robs$. Both half-mass and half-light radii are smaller than the observed radii, which are subject to PSF effects. We provide further details on the size biases across radius definitions in Appendix~\ref{ap:size-bias}. At early times, half-mass sizes are comparable to, or slightly under, half-light sizes, and continue to grow above these as galaxies develop their old stellar population. The slightly larger half-light sizes at early times and the compact galaxies scenario found by the \RTnsCRiMHD model are in agreement with observational data suggesting no inside-out star formation at this epoch \citep{Ono2024CensusFormation}. 

PSF convolution leads to observed sizes that are closer to, but systematically larger than, the EPOCHS results and reduces size scatter. For JWST F444W, the PSF is approximately sampled by 2~pixels. The physical size of the PSF FWHM ranges from 0.52 kpc at $z=12$ to 1.11 kpc at $z = 3$. This largely exceeds the median intrinsic galaxy sizes by up to about a factor of two. Low-mass galaxies are particularly affected, leaving a large proportion of these systems unresolved and making size reconstruction difficult. The gap between observed and half-light sizes narrows at lower redshifts, as galaxies grow more rapidly than the PSF increases and become better resolved. 

This trend toward improved resolution is further supported by the $R\,(z)$ relation measured for more massive and extended galaxies by \citet{Allen2024GalaxyImaging}, and extrapolated here to lower masses. These analyses recover smaller sizes than EPOCHS at higher redshift ($z \gtrsim 5$), with values that align more closely with the intrinsic galaxy sizes (right panel) at high redshift, especially at $z \sim 6.5$.

%%%%%%%%%%% Redshift-r, mass bins %%%%%%%%%%%
\begin{figure*}[ht!]
    \centering
    \includegraphics[width=\textwidth]{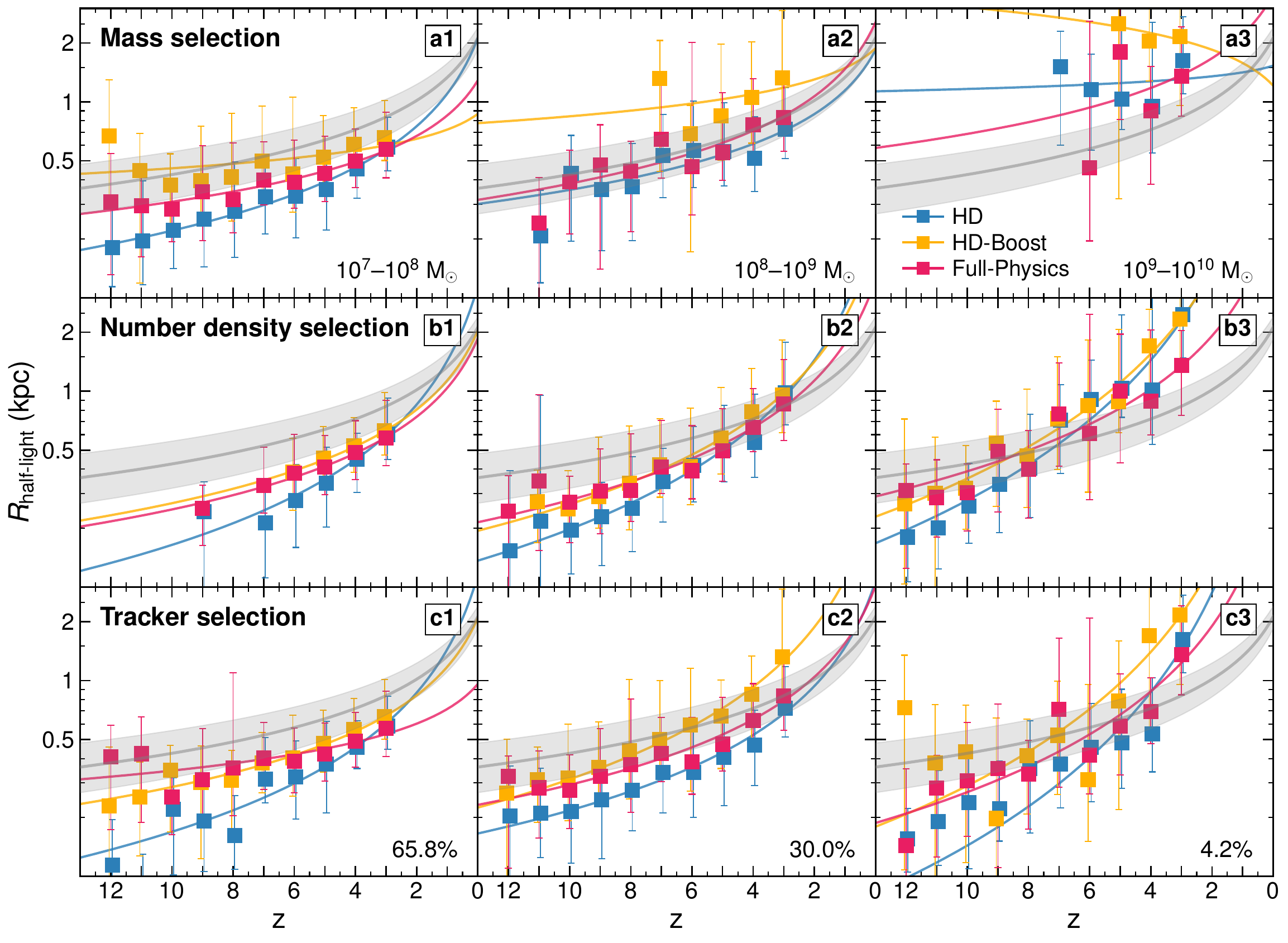}
    \caption{Evolution of the galaxy half-light radii ($\Rhl$) as a function of redshift under different galaxy selections. From left to right, columns display results from lower to higher mass bins. Rows correspond to different galaxy selection methods: fixed stellar mass bins (top), number-density selection (center), and tracker selection (bottom; see text for details). For each panel, we compare results across our three models: standard HD (blue), enhanced SN feedback HD-Boost (yellow), and \RTnsCRiMHD (red). Points show median galaxy sizes, with error bars corresponding to the 16th and 84th percentiles. Lines show the best fitting power-laws for each model. The shaded gray region and line show the EPOCHS results at $7 < z < 12$ \citep{Westcott2025EPOCHS12.5}. In the top row, the bottom right text shows the mass range of the galaxies included in the panel. In the bottom row, the text shows the fraction of galaxies within the panel relative to the full $z=3$ sample for the number density and tracker selections. 
    %
    %Due to sample scarcity at higher redshifts, we restrict the rightmost panel in each column to $z \leq 6$.
    %
    Across mass ranges, the lowest selection bin ($10^7 - 10^8~\text{M}_\odot$) provides the strongest discrimination between galaxy formation models. Sizes evolve with steeper slopes at lower redshifts under tracker and number density selections, suggesting that individual galaxy assembly dominates the evolution of $R\,(z)$ over cosmic time.}
    \label{fig:mass-bin-z-r}
\end{figure*}

To understand how different galaxies contribute to the global size growth of galaxies across redshift, we explore how different galaxy selection criteria affect the growth of galaxies. 
Figure~\ref{fig:mass-bin-z-r} shows the evolution of median half--light radii (i.e., without PSF convolution) as a function of redshift, from $z \sim 12$ to $z \sim 3$. We compare results across our three models: HD, HD-Boost, and \RTnsCRiMHD. 
Rows show different galaxy selection methods: stellar mass bins, number density selection, and tracker selection, respectively. We detail below how these selections are performed. For comparison, the gray line and shaded region indicate the EPOCHS survey radius--redshift relation. Points show median sizes with 16th--84th percentile error bars. Our error bars account for both the measurement uncertainty of median galaxy sizes, and the statistical uncertainty from finite sample sizes, estimated as a Poisson error. 

\textbf{Stellar mass selection (top row, panels a1--a3):} We separate galaxies according to their stellar mass at each given redshift, to investigate how the properties of galaxies in a given selection bin vary across time. Panels correspond to mass bins of $10^{7}–10^{8}~\text{M}_\odot$, $10^{8}–10^{9}~\text{M}_\odot$, and $10^{9}–10^{10}~\text{M}_\odot$. 
This approach is analogous to a stellar mass selection in an observational sample, reflecting how the size of galaxies of a given mass changes across redshift.  

If galaxy populations do not undergo size evolution, trajectories across the $R\,(z)$ space will be flat. Overall, we recover similar trends as above: larger sizes for the HD-Boost model and smaller sizes for the HD model. All models and mass ranges feature large size scatters.
The HD model has a steeper evolution across $z \sim 9–5$ than the other two models, especially for $\Mstar \lesssim 10^{8}\,\Msun$ (panel a1). Such evolution indicates rapid galaxy size growth at late times. Conversely, in the shallower evolution models, the modulation of the global $R\,(z)$ relation is less affected by morphological evolution of galaxy populations.
This serves as another discriminant between models, especially for the $z \gtrsim 5$ range and stellar masses in $\Mstar \in [10^{7},\,10^{8}]\,\Msun$. Notably, this mass range lies near the peak of the EPOCHS data mass distribution. 

\textbf{Number density selection (middle row, panels b1--b3):} The comparatively flatter trajectories in the \RTnsCRiMHD and HD-Boost models under a mass selection suggest that galaxies undergo steeper size evolutions across their lifetimes. 
We investigate a number density selection (middle row), separating galaxies at fixed cumulative number densities across cosmic time. This provides a statistical proxy for tracing the evolution of galaxy mass ranks across redshift. To facilitate comparison with our mass selection method, we compute number density thresholds by mapping directly onto the mass ranges of the top row. The number density thresholds are measured in the \RTnsCRiMHD model at each given redshift, and applied to our three simulations. 
At high redshift, this selection naturally produces sparse sampling in the leftmost panels. This arises from our resolution limits and later turn-around time for smaller halo progenitor perturbations.

The size evolution follows clear power-law relations for all number density bins and models, and also features smaller scatter than the mass selection. Number density selection curves also have steeper slopes than their mass selection counterparts. This indicates significant size growth of galaxies through cosmic time, and suggests that the global increase of $R\,(z)$ across cosmic time is dominated by the contribution from individual galaxies evolving along the size--mass relation (i.e., first term in $\frac{dR}{dz} = \left(\frac{\partial R}{\partial M_*}\right)_z \frac{dM_*}{dz} + \left(\frac{\partial R}{\partial z}\right)_{M_*} $).

Under our number density selection, HD galaxies remain systematically smaller than their HD-Boost and \RTnsCRiMHD counterparts at high redshifts. While HD-Boost still produces the largest sizes, these are comparable to the sizes from the \RTnsCRiMHD model. This result is further explained in our review of the tracker selection (bottom row). Toward lower redshifts, and especially for our most massive galaxies (panel b3), the separation between models decreases. By $z \sim 3$, sizes resulting from different galaxy formation physics no longer follow systematic trends and are comparable within the scatter.

\textbf{Tracker galaxy selection (bottom row, panels c1--c3):}
The number density selection provides a simple and powerful proxy to estimate the growth of galaxy populations. However, the assumption of conserved rank ordering is violated by the large scatter in assembly histories \citep{Behroozi2013The0-8}. 
Instead, we use galaxy trackers, which follow individual galaxies backward through cosmic time and reconstruct their assembly histories.
% We better understand the evolutionary trajectories of galaxies. 
% This is further compounded by satellite galaxies being removed from this relation over redshift, and a higher scatter at early times.
% Monotonic rank ordering
%
We adopt the same three number density bins from the previous selection, assigning all $z=3$ galaxies across models using the thresholds derived from the \RTnsCRiMHD model. The bottom right text shows the fraction of galaxies within each panel relative to the full $z=3$ sample.
Once a galaxy has a tracker selection bin assigned, it remains in that bin across all redshifts. This allows us to track how the main progenitors of galaxies within a given mass range at $z = 3$ evolve across redshift. This method exactly follows the evolutionary tracks of galaxies in our simulation\footnote{Note that by construction, this approach excludes all non-main progenitor systems that merged with other galaxies prior to $z = 3$.}, which is particularly important for interpreting the similarity between HD-Boost and \RTnsCRiMHD under the number density selection.

The overall trends observed for the other two selection methods are preserved for our tracker selection, with a closer similarity to the number density selection.
Despite this, we observe various important differences. In the lowest and intermediate mass-ranked bins (panels c1, c2), \RTnsCRiMHD galaxies have a shallower trajectory, building a higher proportion of their final sizes and masses at earlier times. This trend was already seen for our highest masses in the number density selection, as these are less subject to assembly scatters within our limited volume. In the selection equivalent to the bin $\Mstar \in [10^{7},\,10^{8}]~\Msun$ (panel c1), the physical size variations appear to be set early on by our models, and retained as they evolve in redshift. 

%%%%%%%%%%% SURFACE BRIGHTNESS %%%%%%%%%%%
\subsection{\RTnsCRiMHD model produces detectable population of compact, bright galaxies} \label{subsec:results-surface-brightness}
Galaxy sizes are not only important to study the morphologies of the first galaxies. Size and morphological variations also determine how galaxy brightness is distributed, and ultimately which systems are detectable. 

\begin{figure*}[ht!]
    \centering
    \includegraphics[width=\textwidth]{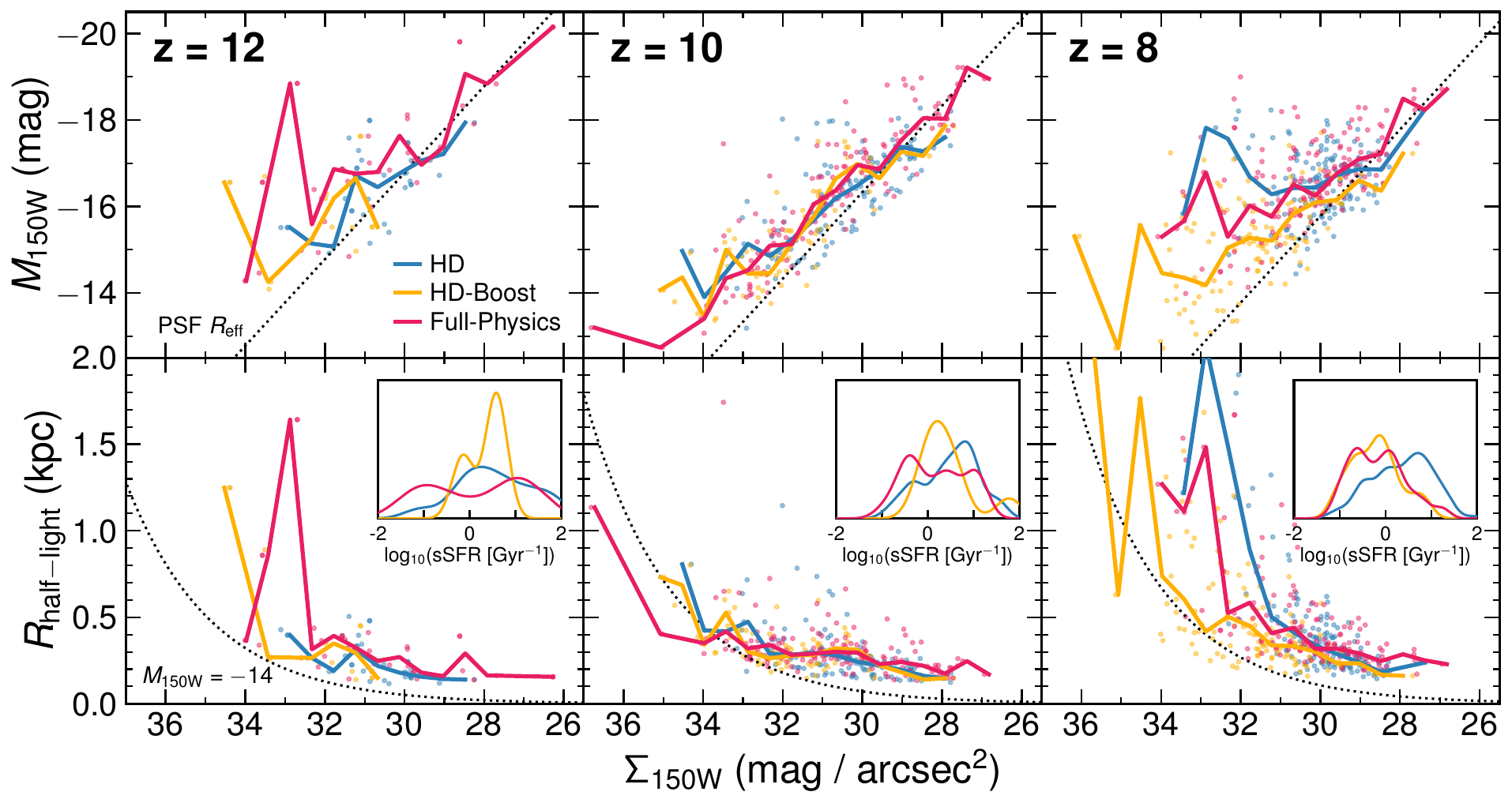}
    \caption{(Top) Absolute UV magnitude versus apparent surface brightness $\Sigma_\text{150W}$ from our synthetic observations. (Bottom) UV half-light radius ($\Rhl$) versus apparent surface brightness. We compare our three models, with each column from left to right progressing from $z = 12$, $10$, to $8$. Solid lines show running medians computed in 20 uniform magnitude bins. Black dotted lines in the top row indicate the variation of absolute magnitude as a function of $\Sigma_\text{150W}$ at a fixed $\Rhl$, set to the F150W PSF at each redshift. Black dotted lines in the bottom row indicate variations of $\Rhl$ as a function of $\Sigma_\text{150W}$, for a fixed absolute UV magnitude of $M_\text{150W} = -14\,\magunits$. 
    Insets in the bottom panels display the specific star formation rate distribution for the $10^7–10^8~\text{M}_\odot$ stellar mass range.
    Across all three redshifts, \RTnsCRiMHD galaxies reach systematically higher and broader surface brightnesses. The brightest galaxies, which are most likely to be detectable, have the smallest sizes and are typically unresolved.}
    \label{fig:surfacebrightness}
\end{figure*}

We show in the top row of Figure~\ref{fig:surfacebrightness} the absolute UV magnitude as a function of the apparent UV surface brightness for our synthetically observed galaxies, now using the JWST F150W filter. Panels compare our results across our three models: HD, HD-Boost, and \RTnsCRiMHD. Columns show results at redshifts $z = 12$, $z = 10$, and $z = 8$. We compute the apparent surface brightness $\Sigma_\text{150W}$ by converting the galaxy's flux to apparent magnitude and adding the logarithmic contribution from its projected area on the sky. This area is estimated using the half-light radius and the angular-diameter distance. We show running medians for each of the models and the variation of the absolute magnitude as a function of the apparent surface brightness, for a fixed UV physical size $\Rhl$. We set this size to the JWST/NIRCam F150W PSF at the redshift shown.

For all redshifts shown, \RTnsCRiMHD galaxies have the broadest distribution of surface brightnesses across models. Its brightest galaxies are systematically brighter than those in HD and HD-Boost, typically reaching $\Sigma_\text{150W} \sim 26–28\,\SigmaUVunits$ values. This is particularly notable at $z = 12$, with \RTnsCRiMHD systems reaching significantly higher surface brightnesses, whereas HD-Boost systems typically have $\Sigma_\text{150W} \sim 33–35\,\SigmaUVunits$ range. At $z = 10$, all three models have roughly similar magnitude--surface brightness slopes. However, \RTnsCRiMHD extends to both brighter and fainter surface brightnesses than HD and HD-Boost. At $z = 8$, the $\SigmaUV$ distribution becomes broader for all models. Furthermore, the scaling of the relation flattens toward dimmer surface brightnesses ($\Sigma_\text{150W} \sim 31–36\,\SigmaUVunits$), driven by size variations. The relations for the three models differ significantly at this faint end, converging at intermediate surface brightness ($\Sigma_\text{150W} \sim 28–32\,\SigmaUVunits$). Toward the bright end ($\Sigma_\text{150W} \sim 26–28\,\SigmaUVunits$), the relations diverge again and steepen with different behaviors. 

To deconvolve the contributions from absolute magnitude and galaxy sizes, we show in the bottom row of Figure~\ref{fig:surfacebrightness} the UV half-light radius as a function of the apparent surface brightness. We show the variation of $\Rhl$ as a function of apparent surface brightness for a fixed absolute UV magnitude, set to $M_\text{150W} = -14~\magunits$, corresponding to relatively faint galaxies. 
We include in the top right of each panel the specific star formation rate\footnote{We compute the star formation rate over the past 100 Myr.} (${\rm sSFR} \equiv {\rm SFR}/\Mstar$) for the $10^7–10^8~\Msun$ stellar mass range, using a kernel density estimation fit. 

Across all redshifts shown, our models show a large size scatter at low surface brightnesses. Their relations converge to a flatter slope with lower scatter at $\Sigma_\text{150W} \lesssim 31\,\SigmaUVunits$. This result suggests the existence of a population of large and low surface brightness galaxies below current detection thresholds. Our models also predict a population of highly compact, actively star-forming bright galaxies. Their small sizes and high brightness make these systems more likely to be detected by JWST observations, but also more likely to be unresolved. 
Overall, all of our models predict the high-redshift galaxy population detected by JWST is dominated by bright, compact, and unresolved systems. 

\begin{figure}[ht!]
    \centering
    \includegraphics[width=\columnwidth]{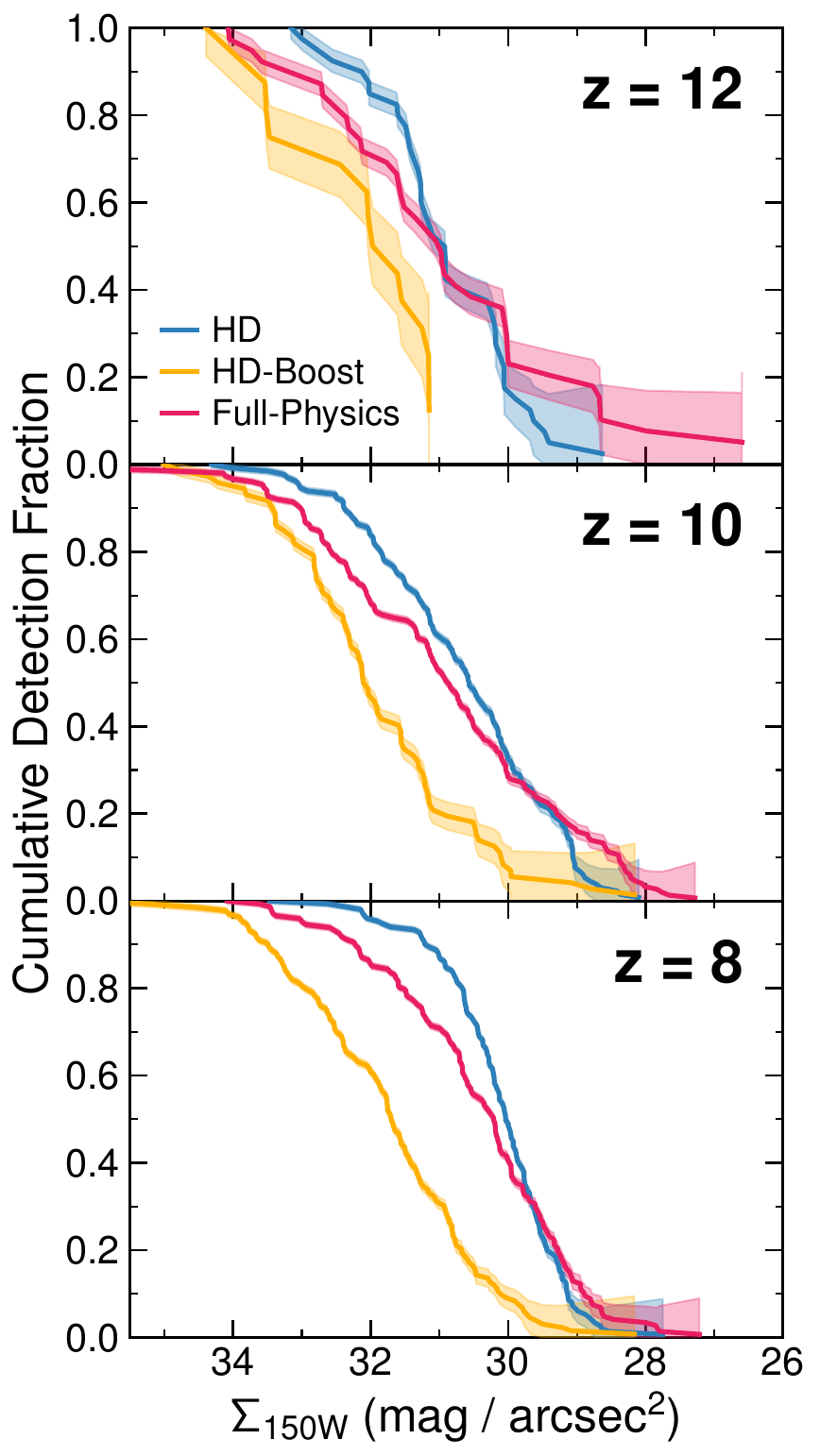}
    \caption{Cumulative distribution function of detectable galaxies as a function of peak apparent surface brightness detection threshold. From top to bottom, panels show our three physical models at $z = 12$, $z = 10$, and $z= 8$. 
    At these redshifts, \RTnsCRiMHD galaxies are more likely to be detected by JWST due to their higher surface brightnesses.}
    \label{fig:detection-fraction}
\end{figure}

The differences in surface brightness across our models lead to variations in galaxy detectability. With galaxy sizes being particularly small toward the bright end, we approximate the detectability of galaxies using the surface brightness of these objects, estimated using $\sqrt{\Rhl^2 + \text{R}_\text{PSF, eff}^2}$, where $\text{R}_\text{PSF, eff}$ is the PSF physical size. 
% NOTE: no nebular line emission contribution to UV filters is incorporated. This can make our galaxies appear fainter, especially at high sSFR. 
%
Figure~\ref{fig:detection-fraction} shows the cumulative detection fraction of galaxies across our models as a function of apparent surface brightness detection threshold, $\SigmaUV$, for $z = 12$, $z = 10$, and $z = 8$, comparing our HD, HD-Boost, and \RTnsCRiMHD models.

The HD and HD-Boost models have comparably steep slopes across the bulk of their population, only displaced by $\sim 2\,\SigmaUVunits$. This indicates that most of their galaxies are clustered within $\Delta\SigmaUV \sim 3\,\SigmaUVunits$. Conversely, the \RTnsCRiMHD model produces a substantially broader distribution ($\Delta\SigmaUV \sim 6\,\SigmaUVunits$), with galaxies as faint as those in the HD-Boost model and extending to brighter systems than those found in the standard HD case. As a result of its broader surface brightness distribution, the \RTnsCRiMHD model dominates the fraction of galaxies detectable at shallower brightness limits. This results from the combination of smaller galaxy sizes and bursty star formation, shown in the sSFR distribution in Figure~\ref{fig:surfacebrightness}.

The differences between our models are the most pronounced at $z = 12$. While all three models concentrate their population at roughly $\Sigma_\text{150W} \sim 31\,\SigmaUVunits$, \RTnsCRiMHD galaxies dominate the brightest end of the distribution ($\Sigma_\text{150W} \sim 26.5\,\SigmaUVunits$). For comparison, the detection limit computed from the JADES Deep GS F150W depth with a 0.16\arcsec\ radius aperture is $\sim27~\SigmaUVunits$ \citep{Conselice2025EPOCHS.Data}, and $\sim25~\SigmaUVunits$ for CEERS with a 0.1\arcsec\ radius aperture \citep{Finkelstein2023CEERSJWST}. At this redshift, the sSFR distribution of the \RTnsCRiMHD model is the broadest (Figure~\ref{fig:surfacebrightness}), with a bimodal population of star-forming and quiescent galaxies. This requires galaxies to concentrate their star formation in short periods, suggesting a preference for bursty star formation histories. Instead, the HD and HD-Boost samples are dominated by star-forming galaxies.  

As redshift decreases, differences between model sizes progressively decrease. By $z = 10$, the \RTnsCRiMHD model maintains a broad surface brightness distribution, still with a higher bright end. The HD model builds up a larger population of bright galaxies, as star formation proceeds unregulated in this model. HD-Boost galaxies remain less detectable at significantly lower surface brightness values. Despite this, the burstier star formation of the HD-Boost model leads to an extended tail toward higher brightness, similar to the \RTnsCRiMHD case. By $z = 8$, HD and \RTnsCRiMHD sizes converge at the brightest end ($\Sigma_\text{150W} \sim 27.5\,\SigmaUVunits$). 

Altogether, our \RTnsCRiMHD model provides a natural explanation for the detection of unexpectedly bright galaxies at high-redshifts. This model shows how these compact and starburst galaxies represent the high-surface-brightness tail of a galaxy population with a broad distribution of surface brightnesses. 

\section{Conclusion} \label{sec:conclusion}
In this work, we investigate how different physical processes shape the morphology of galaxies and their size evolution from $z = 12$ to $z = 3$, using the high-resolution cosmological zoom-in \textsc{Azahar} simulation suite. We compare the evolution of thousands of resolved galaxies using three physical models: one with standard hydrodynamics (HD), one with enhanced SN feedback (HD-Boost), and a comprehensive model incorporating radiative transfer, cosmic rays, and magnetic fields (\RTnsCRiMHD). 

Overall, we find that more comprehensive physical models for galaxy formation{---}through the inclusion of radiative transfer, cosmic rays, and magnetic fields{---}produce systematically different morphologies than simpler models, especially those relying on calibrated supernovae feedback.

Our main findings can be summarized as follows:
\begin{enumerate}
    \item Treatment of galaxy formation physics leads to observable differences in the morphological appearance of galaxies. Our comprehensive \RTnsCRiMHD model produces structured galaxies with extended disks, well-defined dust patterns, and distributed star formation. Calibrating SN feedback strength to the galaxy stellar mass function at $z = 3$ leads to diffuse, fragmented, and unstructured galaxies, whereas standard HD without calibration leads to compact, highly concentrated galaxies. 
    \item At high stellar masses ($\Mstar \gtrsim 10^{8}\,\Msun$), the \RTnsCRiMHD model produces regulated sizes that are in excellent agreement with observed sizes. This agreement persists when extrapolating the size--mass relation to lower masses ($\Mstar = 10^7$--$10^{8}~\Msun$). In contrast, the HD-Boost model produces $\Rhl$ galaxy sizes that are nearly twice as large as observations.
    \item At low stellar masses ($\Mstar \lesssim 10^{8}\,\Msun$), inferred galaxy sizes reported by observations are significantly smaller than the instrumental resolution (e.g., $R_e \sim 400\,\pc$, approximately half the PSF size). These measurements represent PSF-dominated upper limits and deviate from the more compact sizes extrapolated from higher-mass galaxy observations. 
    \item We find that number-density selection preserves systematic, per-model trends and offers a practical avenue to follow galaxy size assembly, but individual galaxy tracks show steeper size growth. This indicates that individual galaxy assembly dominates size evolution at early times.
\end{enumerate}

Our \RTnsCRiMHD model naturally explains the abundance of bright galaxies found by JWST within  $\Lambda\text{CDM}$, and makes a clear prediction: the galaxies observed by JWST at the highest redshift are the bright, compact, star-bursting end of a broader distribution. This indicates that there is a significant population of fainter and lower surface brightness galaxies yet to be uncovered at these times.

We highlight the importance of better characterizing observational galaxy sizes in the $\Mstar \in \left[10^{7},\, 10^{8}\right]\,\Msun$ regime, which we find are most sensitive to galaxy formation models. At $z \gtrsim 10$, the models in this mass regime vary by a factor of 3 and predict sizes between $\Rhl = 200–500~\pc$. However, these predicted galaxy sizes are below JWST resolution limits, making accurate size reconstruction difficult. As galaxies grow larger and become better resolved, intrinsic and observed sizes converge. Real galaxy effective radii may be even smaller than our predictions, if their structure concentrates at scales below our spatial resolution ($\sim 20~\pc$). 

Our high-redshift galaxy size predictions will be testable with the next generation of Extremely Large Telescopes, which should be able to resolve compact galaxies down to sizes of $\sim 100~\pc$ at $z \sim 10$ (e.g., with $R_e / R_\text{PSF} \sim 3$--7, or $>10\times$ resolution improvement; \citealt{Davies2018TheSimulation}). Such observations could then directly test the physical origin of the galaxy population emerging at cosmic dawn. 

% Beyond resolving the first galaxies, another key test is to trace their evolution to the present day. We propose tracing the first galaxies in the $\Mstar \in \left[10^{7},\, 10^{8}\right]\,\Msun$ regime to $z = 0$ to answer whether the power-law growth seen from $z = 12$ to $z = 3$ persists through the $z = 3$ to $z = 0$ epoch, or if galaxy growth mechanisms transition after the Cosmic Noon. Abundance matching our $z \sim 3$ galaxies to $z = 0$ would identify the observable, well-resolved descendants of our high-redshift population at $0 \leq z \leq 3$, enabling low-redshift surveys to test whether growth mechanisms at $z>3$ persist beyond the Cosmic Noon.

\begin{acknowledgments}
We thank the anonymous referee for reading the manuscript and providing useful feedback. This work received support from the Kavli Institute for Particle Astrophysics and Cosmology and the Vice Provost for Undergraduate Education at Stanford University.

This work used the DiRAC@Durham facility managed by the Institute for Computational Cosmology on behalf of the STFC DiRAC HPC Facility (www.dirac.ac.uk). The equipment was funded by BEIS capital funding via STFC capital grants ST/P002293/1, ST/R002371/1 and ST/S002502/1, Durham University and STFC operations grant ST/R000832/1. DiRAC is part of the National e-Infrastructure. 
Some of the computing for this project was performed on the Sherlock cluster at Stanford. We would like to thank Stanford University and the Stanford Research Computing Center for providing computational resources and support that contributed to these research results. 
\end{acknowledgments}

%\begin{contribution}
%\end{contribution}

\appendix
\counterwithin{figure}{section}   

\section{Assumption of Gaussian PSF} \label{ap:psf-shape}
In this appendix we review the validity of the Gaussian PSF assumption by comparing it with a realistic JWST PSF. The JWST PSF is shaped by its hexagonal primary mirror, which produces a central core and diffraction spikes. We construct this PSF from the Fraunhofer diffraction pattern of a hexagonal aperture scaled so that the resulting PSF FWHM matches the F444W FWHM  at each redshift.

\begin{figure}[ht!]
    \centering
    \includegraphics[width=\columnwidth]{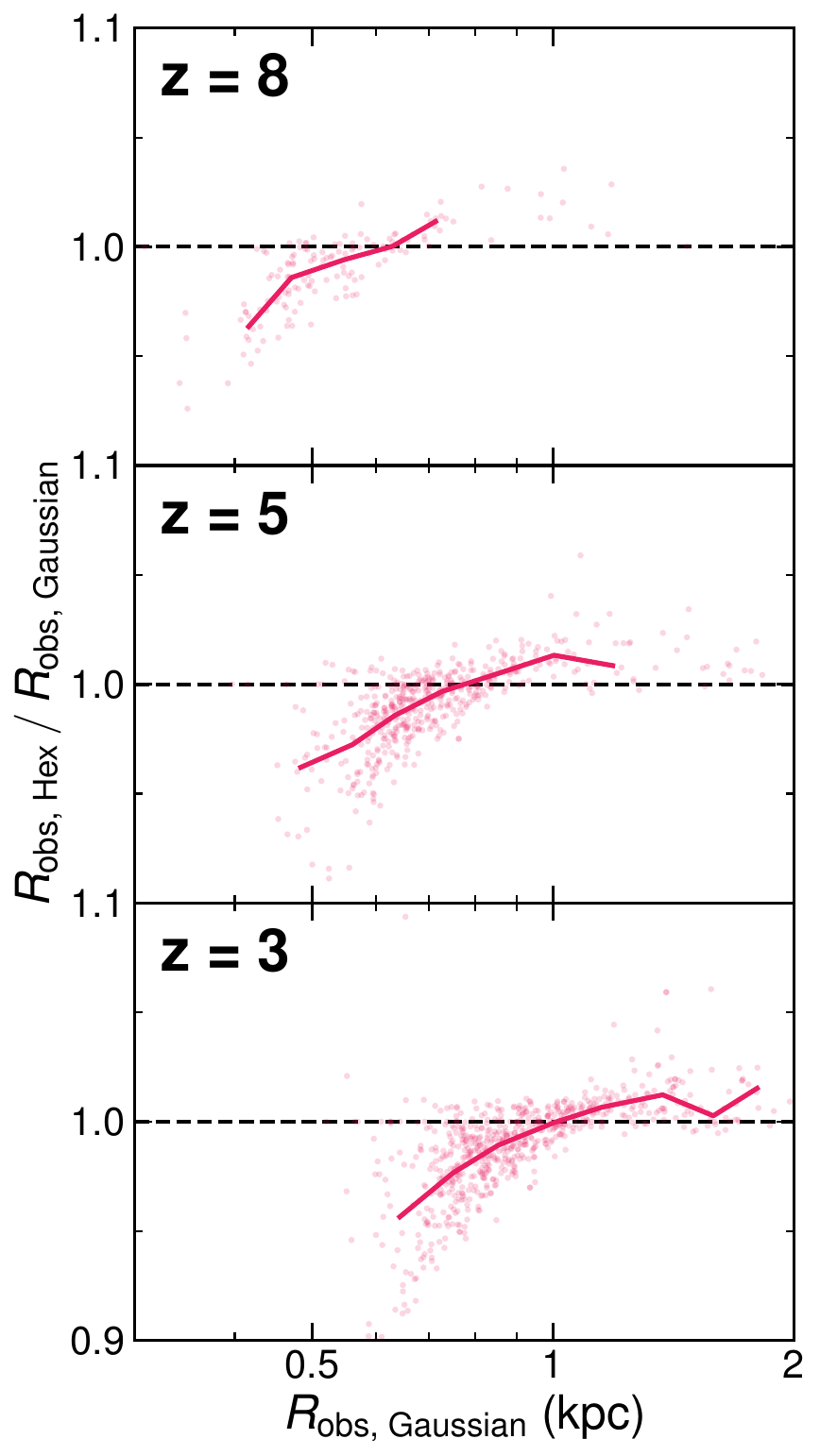}
    \caption{Ratio of observed sizes from the \RTnsCRiMHD model after convolution with a realistic JWST hex-aperture PSF ($\mathrm{R}_{\mathrm{obs,\,Hex}}$) to those with a Gaussian PSF ($\mathrm{R}_{\mathrm{obs,\,Gaussian}}$). From left to right, panels show redshifts $z=8$, $5$, and $3$. In each panel, the light points are individual galaxies and the solid lines show the running medians in 15 log-spaced bins across $0.2$--$2.0~\kpc$ with more than 10 galaxies. The horizontal dashed line marks the 1:1 ratio. Across our full size range, the two PSF shapes yield observed sizes that agree to within $\lesssim 5\%$.}
    \label{fig:psf-shape-comparison}
\end{figure}

Figure~\ref{fig:psf-shape-comparison} shows the ratio $\mathrm{R}_{\mathrm{obs,\,Hex}} /  \mathrm{R}_{\mathrm{obs,\,Gaussian}}$ as a function of $\mathrm{R}_{\mathrm{obs,\,Gaussian}}$ for the \RTnsCRiMHD model at $z = 8$, $5$, and $3$. The dashed line indicates a $1:1$ ratio where the two PSFs yield identical observed sizes.

The resulting ratios are centered on 1 across all redshifts, with deviations below $5\%$ across the full size ranges. At the smallest sizes ($\mathrm{R}_{\mathrm{obs,\,Gaussian}} \lesssim 0.7 ~\kpc$ at $z = 3$), the hex PSF yields slightly smaller sizes than the Gaussian PSF because its core concentrates more flux at the center. Therefore, the assumption of a Gaussian PSF is sufficient for the size measurements in this work.

\section{Fitting results for size--mass relations} \label{ap:size-mass-fits}

\begin{figure*}[h]
    \centering
    \includegraphics[width=0.495\textwidth]{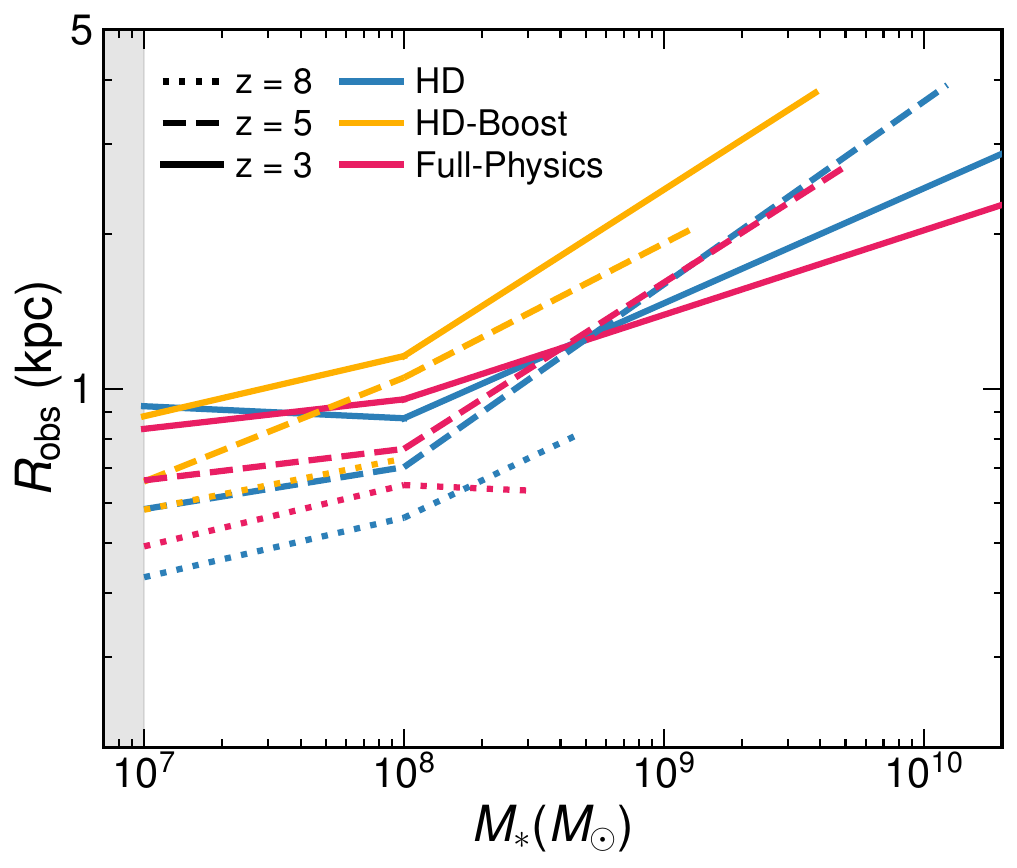}
    \includegraphics[width=0.495\textwidth]{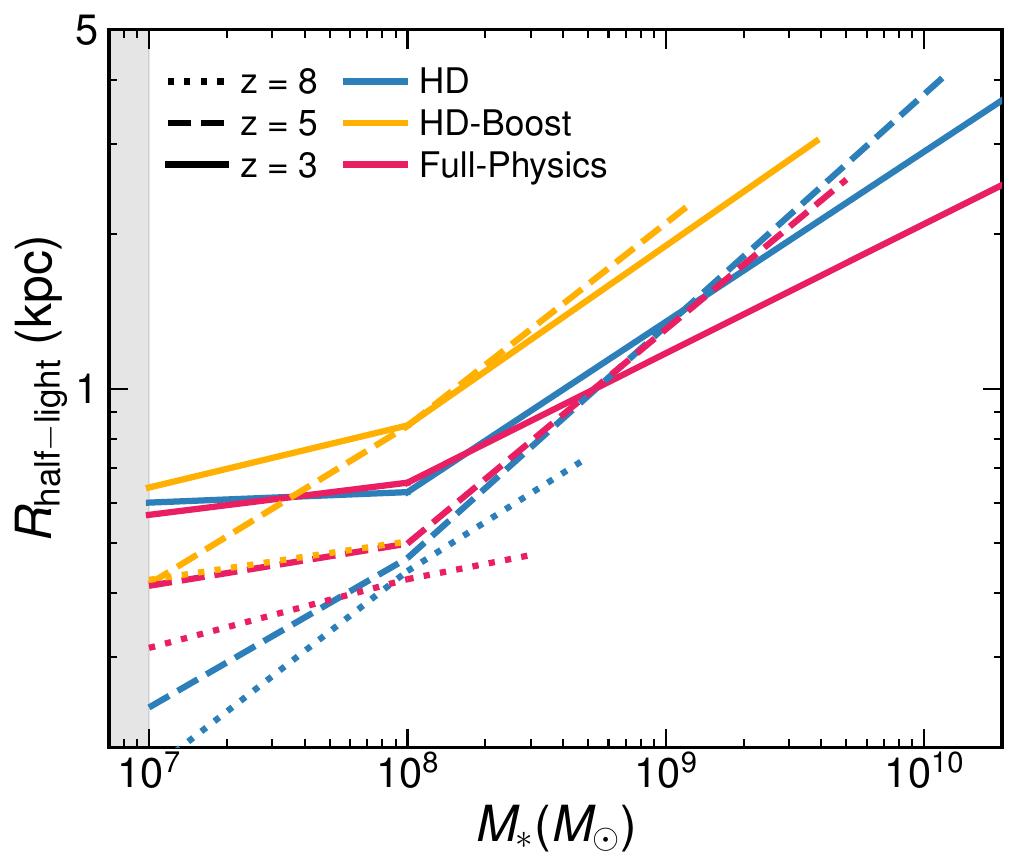}
    \caption{Size--mass relations for our two observational (JWST 444W) size definitions: (Left) PSF-convolved half-light radius $\Robs$, and (right) half-light radius prior to PSF convolution $\Rhl$. Each panel shows the best-fit double power law relations for our three studied models: HD in blue, HD-Boost in yellow, and \RTnsCRiMHD in red. The line styles represent the redshift: dotted for $z = 8$, dashed for $z = 5$, solid for $z = 3$. The gray shaded band indicates galaxies below $\Mstar < 10^7~\Msun$, which are less resolved.} 
    \label{fig:size-mass-fits}
\end{figure*}

%% NEXT APPENDICES
\renewcommand{\thefigure}{C.\arabic{figure}}
\setcounter{figure}{0}
\makeatletter
\renewcommand{\theHfigure}{C.\arabic{figure}}
\makeatother
\begin{figure*}[h]
    \centering
    \includegraphics[width=\textwidth]{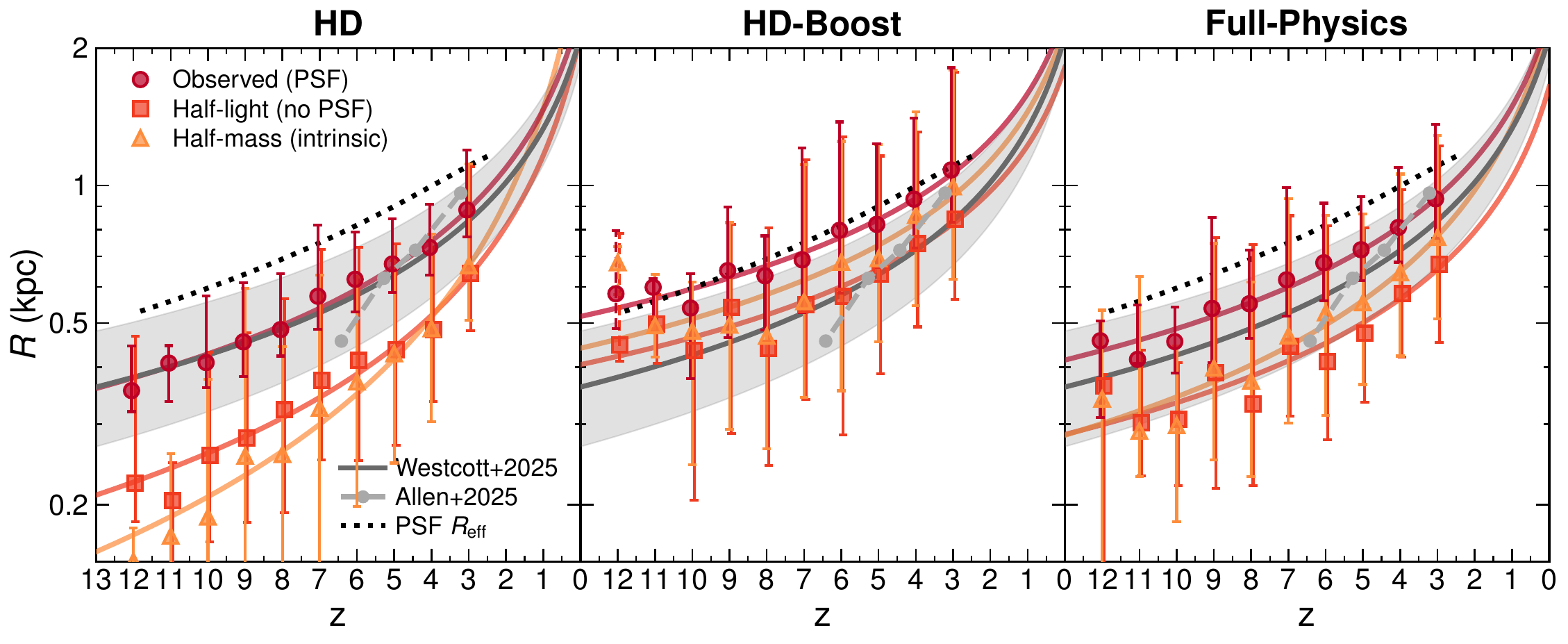}
    \caption{Same as the right panel of Figure~\ref{fig:z-r-with-literature}, now for each of our three studied models. From left to right, we show HD, HD-Boost, and \RTnsCRiMHD. Error bars indicate the 16th and 84th percentiles.}
    \label{fig:z-r-fits}
\end{figure*}

Figure~\ref{fig:size-mass-fits} shows the size--mass relation fits for our two synthetic observation size definitions. The left panel shows the PSF-convolved half-light radius in JWST 444W ($\Robs$) and the right panel shows the half-light radius prior to PSF convolution ($\Rhl$).
Following Figure~\ref{fig:size-mass-evol}, each panel compares the three studied models: HD, HD-Boost, and \RTnsCRiMHD. We show results for $z = 8$, $z = 5$, and $z = 3$. 

We find evidence for a break in the size--mass relation at $\Mstar = 10^8~\Msun$, consistent with previous studies \citep{Chamba2024AGalaxies, McClymont2025TheUniverse}, and supported by our error analysis. We restrict our fitting of the size--mass relation for galaxies with $\Mstar > 10^7~\Msun$, which is our stellar mass low resolution limit. As a result, we fit a double power-law, with a break fixed at  $\Mstar = 10^8~\Msun$ and enforce continuity at the break mass: 

\iftrue
\begin{equation}
% better compact version
\log(R/\text{kpc}) = \log A_i + \alpha_i \log(\Mstar/\Msun),
\label{eq:size_mass}
\end{equation}
with $i=1$ for $10^7 \leq \Mstar/\Msun < 10^8$ and $i=2$ for $\Mstar/\Msun \geq 10^8$. %
\else
\begin{align}
\log(R/\text{kpc}) &= \log(A_1) + \alpha_1 \log(\Mstar / \Msun) \quad \text{for } 10^7 \leq \Mstar < 10^8~\Msun, \\
\log(R/\text{kpc}) &= \log(A_2) + \alpha_2 \log(\Mstar / \Msun) \quad \text{for } \Mstar \geq 10^8~\Msun.
\label{eq:size_mass}
\end{align}
\fi
In Table~\ref{tab:s-m-fits} we provide the best-fit parameters and their errors for the size--mass relation across $3 \leq z < 12$, for $\Robs$ (left) and $\Rhl$ (right). For each mass range, we report the slope $\alpha$, intercept $\log(A)$, and intrinsic scatter $\sigma\,\log(R)$, defined as the distribution in galaxy sizes. Entries marked as ``--'' indicate bins containing fewer than 10 galaxies and are therefore not reported.

\begin{table*}[h]
\centering
\caption{Best-fit parameters to the size--mass relation for (left) $R_{\text{obs}}$ and (right) $\Rhl$.}
\hspace{-2cm}
\begin{minipage}{0.4\textwidth}
\centering
\setlength{\tabcolsep}{3pt}
\begin{tabular}{c c c c c c c}
\hline
\hline
 & \multicolumn{6}{c}{\large \textbf{Observed size $R_{\text{obs}}$}} \\
\cline{1-7}
$z$
& $\alpha_1$
& $\log(A_1)$
& $\sigma\,\log(R)$
& $\alpha_2$
& $\log(A_2)$
& $\sigma\,\log(R)$ \\
\hline
\multicolumn{7}{c}{\textbf{HD}} \\
\hline
3  & -0.023 &  0.130 & 0.122 &  0.224 & -1.848 & 0.154 \\
4  &  0.044 & -0.474 & 0.087 &  0.233 & -1.990 & 0.149 \\
5  &  0.082 & -0.806 & 0.081 &  0.356 & -3.000 & 0.148 \\
6  &  0.091 & -0.911 & 0.085 &  0.272 & -2.361 & 0.131 \\
7  &  0.071 & -0.788 & 0.095 &  0.260 & -2.305 & 0.123 \\
8  &  0.116 & -1.177 & 0.098 &  0.242 & -2.186 & 0.129 \\
9  &  0.110 & -1.147 & 0.125 &  0.264 & -2.378 & 0.174 \\
10 &  0.035 & -0.660 & 0.072 & -- &  -- & -- \\
11 & -0.436 &  2.799 & 0.158 & -- &  -- & -- \\
12 &  0.066 & -0.936 & 0.062 &   --  &   --   &  --   \\
\hline
\multicolumn{7}{c}{\textbf{HD-Boost}} \\
\hline
3  &  0.117 & -0.874 & 0.135 &  0.324 & -2.528 & 0.201 \\
4  &  0.193 & -1.488 & 0.122 &  0.279 & -2.175 & 0.188 \\
5  &  0.202 & -1.597 & 0.106 &  0.262 & -2.070 & 0.245 \\
6  &  0.328 & -2.530 & 0.151 &  -0.244 & 2.045 & 0.140 \\
7  &  0.226 & -1.810 & 0.156 &  -- & -- & -- \\
8  &  0.100 & -0.933 & 0.114 &  -- & -- & -- \\
9  &  0.168 & -1.447 & 0.151 &   --  &   --   &  --   \\
10 & -0.591 &  3.891 & 0.457 &   --  &   --   &  --   \\
11 &  3.771 & -28.146 & 0.935 &   --  &   --   &  --   \\
12 &  -- & -- & -- &   --  &   --   &  --   \\
\hline
\multicolumn{7}{c}{\textbf{\RTnsCRiMHD}} \\
\hline
3  &  0.058 & -0.481 & 0.118 &  0.165 & -1.337 & 0.200 \\
4  &  0.078 & -0.698 & 0.112 &  0.214 & -1.789 & 0.188 \\
5  &  0.061 & -0.608 & 0.096 &  0.324 & -2.707 & 0.151 \\
6  &  0.141 & -1.240 & 0.102 &  0.260 & -2.193 & 0.213 \\
7  &  0.132 & -1.199 & 0.108 &  0.207 & -1.795 & 0.188 \\
8  &  0.119 & -1.142 & 0.123 &  -- & -- & -- \\
9  &  0.177 & -1.592 & 0.127 & -- &  -- & -- \\
10 &  0.109 & -1.154 & 0.102 & -- & -- & -- \\
11 & -0.033 & -0.094 & 0.162 & -- & -- & -- \\
12 &  0.119 & -1.221 & 0.186 &   --  &   --   &  --   \\
\hline
\hline
\end{tabular}
\end{minipage}
\hspace{0.01\textwidth}
\begin{minipage}{0.4\textwidth}
\setlength{\tabcolsep}{3pt}
\begin{tabular}{c c c c c c c}
\hline
\hline
 & \multicolumn{6}{c}{\large \textbf{Half-light size $\Rhl$}} \\
\cline{1-7}
$z$
& $\alpha_1$
& $\log(A_1)$
& $\sigma\,\log(R)$
& $\alpha_2$
& $\log(A_2)$
& $\sigma\,\log(R)$ \\
\hline
\multicolumn{7}{c}{\textbf{HD}} \\
\hline
3  & 0.020 & -0.364 & 0.211 & 0.332 & -2.853 & 0.198 \\
4  & 0.087 & -1.006 & 0.174 & 0.325 & -2.914 & 0.225 \\
5  & 0.291 & -2.655 & 0.218 & 0.452 & -3.942 & 0.225 \\
6  & 0.306 & -2.793 & 0.240 & 0.329 & -2.976 & 0.233 \\
7  & 0.161 & -1.699 & 0.208 & 0.355 & -3.250 & 0.193 \\
8  & 0.391 & -3.486 & 0.247 & 0.317 & -2.892 & 0.252 \\
9  & 0.240 & -2.372 & 0.270 & 0.405 & -3.694 & 0.277  \\
10 & 0.045 & -0.997 & 0.176 & -- & -- & -- \\
11 & -0.795 & 5.164 & 0.282 & -- & -- & -- \\
12 & 0.316 & -3.033 & 0.175 & --    & --     & --    \\
\hline
\multicolumn{7}{c}{\textbf{HD-Boost}} \\
\hline
3  & 0.121 & -1.038 & 0.211 & 0.350 & -2.870 & 0.237 \\
4  & 0.289 & -2.327 & 0.176 & 0.324 & -2.601 & 0.234 \\
5  & 0.308 & -2.533 & 0.221 & 0.395 & -3.232 & 0.305 \\
6  & 0.520 & -4.133 & 0.304 & 0.505 & -4.018 & 0.273 \\
7  & 0.275 & -2.290 & 0.286 & -- & -- & -- \\
8  & 0.075 & -0.896 & 0.289 & -- & -- & -- \\
9  & 0.249 & -2.187 & 0.288 & --    & --     & --    \\
10 & -0.349 & 1.999 & 0.481 & --    & --     & --    \\
11 & 3.802 & -28.429 & 0.949 & --    & --     & --    \\
12 & -- & -- & -- & --    & --     & --    \\
\hline
\multicolumn{7}{c}{\textbf{\RTnsCRiMHD}} \\
\hline
3  & 0.063 & -0.684 & 0.207 & 0.252 & -2.198 & 0.268 \\
4  & 0.119 & -1.176 & 0.191 & 0.252 & -2.238 & 0.244 \\
5  & 0.082 & -0.958 & 0.215 & 0.417 & -3.640 & 0.240 \\
6  & 0.127 & -1.324 & 0.216 & 0.338 & -3.009 & 0.342 \\
7  & 0.177 & -1.695 & 0.185 & 0.234 & -2.148 & 0.263 \\
8  & 0.134 & -1.441 & 0.274 & -- & -- & -- \\
9  & 0.286 & -2.573 & 0.259 & -- & -- & -- \\
10 & 0.090 & -1.200 & 0.181 & -- & -- & -- \\
11 & 0.072 & -1.043 & 0.267 & -- & -- & -- \\
12 & 0.009 & -0.587 & 0.331 & --    & --     & --    \\
\hline
\hline
\end{tabular}
\end{minipage}
\label{tab:s-m-fits}
\end{table*}

\section{Fitting results for size--redshift relations} \label{ap:z-r-fits}
Figure~\ref{fig:z-r-fits} shows the same information as the right-hand panel of Figure~\ref{fig:z-r-with-literature}, now for our three models. Specifically, the figure shows the three studied size definitions: intrinsic half-mass radius $\Rstellar$, half-light radius prior to PSF convolution $\Rhl$, and the PSF-convolved half-light radius $\Robs$. The error bars for these power-law relations are determined by the $16^\text{th}$ and $84^\text{th}$ percentiles of the size distributions and capped at $\pm3\sigma$, following the same method as \citet{Westcott2025EPOCHS12.5}. The black dotted line shows the physical full width at half maximum size of the JWST/NIRCam F444W PSF.

Across all redshifts and all three models, the observed sizes are systematically the largest, while the half-light radii remain slightly larger than, but comparable to half-mass radii. All $\Rhl$ sizes for HD and \RTnsCRiMHD lie below the physical PSF width, indicating that most galaxies are unresolved. 

The ratio between the sizes before and after PSF convolution is the largest for the HD model. For this simulation, $\Robs$ reaches up to a factor of 2 compared to $\Rhl$ at $z \geq 10$. This effect is reduced in the HD-Boost and \RTnsCRiMHD models, where galaxies are larger and less concentrated. These trends are in agreement with our discussion in Figure~\ref{fig:z-r-with-literature} and reinforce that PSF effects can dominate observed size evolution, especially for compact systems and at high redshift. 

Table~\ref{tab:z-r-fits} presents the best-fit parameters and their errors for the size--redshift relations. We express these relations as power-laws of the form $R = R_0(1+\text{z})^\beta$, following \citet{Shibuya2016MORPHOLOGIESGALAXIES}. 

\begin{table*}[h]
\caption{Best-fit parameters $R_0$ and $\beta$ for the size--redshift relation $R = R_0(1+\text{z})^\beta$. Rows show models (HD, HD-Boost, \RTnsCRiMHD) and column groups show size definitions ($\Robs$, $\Rhl$, $\Rstellar$). }
\tablewidth{0pt}
\centering
\hspace*{-2cm}
\begin{tabular}{ccccccc}
\hline\hline
           & \multicolumn{2}{c}{$\Robs$} & \multicolumn{2}{c}{$\Rhl$} & \multicolumn{2}{c}{$\Rstellar$} \\
Model      & $R_0~(\kpc)$       & $\beta$            & $R_0~(\kpc)$        & $\beta$             & $R_0~(\kpc)$        & $\beta$            \\ \hline
HD         & $2.42 \pm 6.60$    & $-0.72 \pm 1.43$   & $1.82 \pm 7.30$     & $-0.78 \pm 2.12$    & $3.27 \pm 14.36$    & $-1.15 \pm 2.46$   \\
HD-Boost   & $2.37 \pm 5.08$    & $-0.58 \pm 1.10$   & $1.92 \pm 5.31$     & $-0.60 \pm 1.43$    & $2.22 \pm 5.41$     & $-0.61 \pm 1.26$   \\
\RTnsCRiMHD & $2.35 \pm 5.86$    & $-0.66 \pm 1.30$   & $1.48 \pm 5.29$     & $-0.60 \pm 1.84$    & $2.36 \pm 7.62$     & $-0.80 \pm 1.72$  
\end{tabular}
\label{tab:z-r-fits}
\end{table*}

\section{Mass-weighting to the EPOCHS sample} \label{ap:epochs-matching}
To illustrate the effect of the EPOCHS mass-weighting, we show in Figure~\ref{fig:epochs-weighting} the observed size evolution across time for the weighted (solid lines) and unweighted (dashed lines) cases. The black dotted line shows the physical full width at half maximum size of the JWST/NIRCam F444W PSF.

The weighted and unweighted results converge by $z = 3$ as the model mass distribution approaches that of the EPOCHS sample. Differences are most pronounced at higher redshift and for HD-Boost, reflecting our limited volume and the delayed stellar mass growth of the model with the stronger SN feedback.

\renewcommand{\thefigure}{D.\arabic{figure}}
\setcounter{figure}{0}
\makeatletter
\renewcommand{\theHfigure}{D.\arabic{figure}}
\makeatother
\begin{figure}[h!]
    \centering
    \includegraphics[width=\columnwidth]{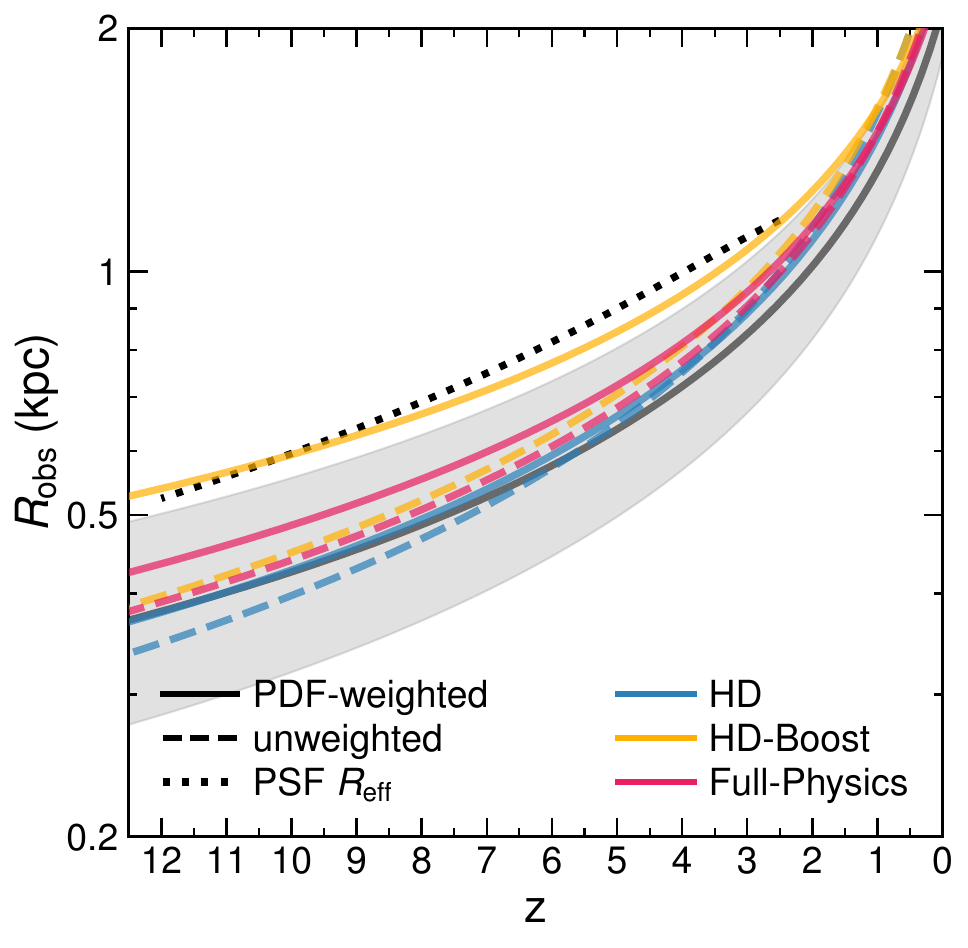}
    \caption{Best-fit power laws for the size evolution of the three physical models: HD, HD-Boost, and \RTnsCRiMHD. Lines compare measurements with (solid lines) and without (dashed lines) weighting our model size distributions to the EPOCHS mass distribution.
    Weighted and unweighted results converge by $z = 3$.}
    \label{fig:epochs-weighting}
\end{figure}

\section{Size variations} \label{ap:size-bias}
\renewcommand{\thefigure}{E.\arabic{figure}}
\setcounter{figure}{0}
\makeatletter
\renewcommand{\theHfigure}{E.\arabic{figure}}
\makeatother
\begin{figure*}[ht!]
    \centering
    \includegraphics[width=\textwidth]{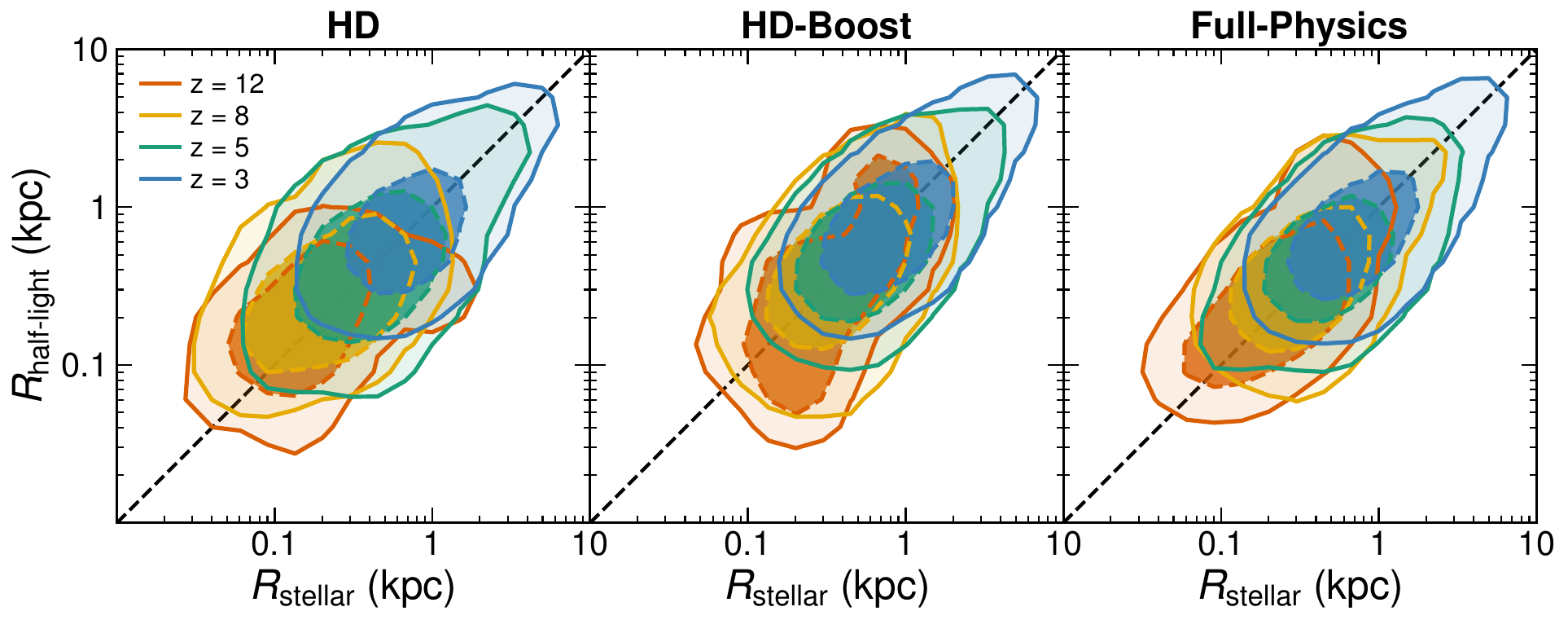}\\
    \includegraphics[width=\textwidth]{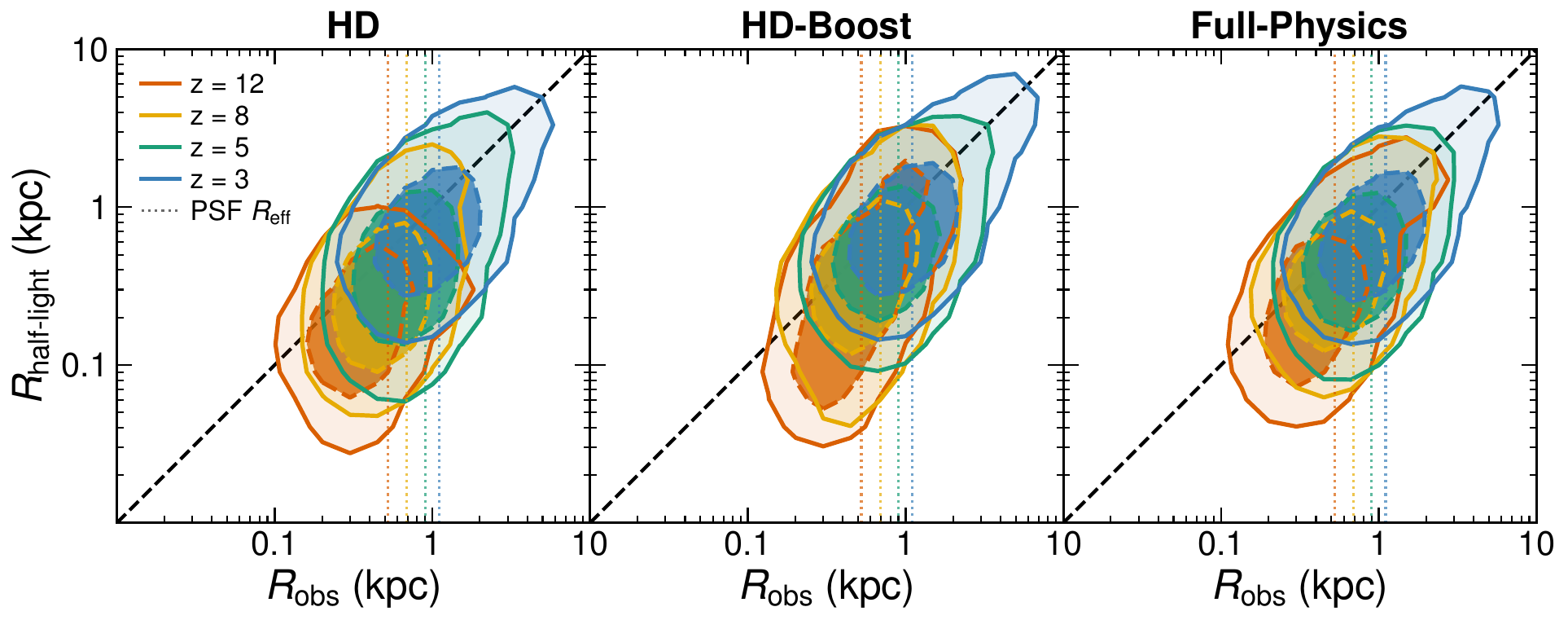}
    \caption{Comparison between: intrinsic size $\Rstellar$ and half-light size $\Rhl$ (top row), and observed size $\Robs$ and half-light size $\Rhl$ (bottom row). 
    From left to right, panels show the three physical models: HD, HD-Boost, and \RTnsCRiMHD. Each panel shows $1\sigma$ (dashed) and $2\sigma$ (solid) kernel density estimate (KDE) contours colored by redshift from $z = 12$ (red), $z = 8$ (orange), $z = 5$ (green), to $z = 3$ (blue). The black dashed line indicates the 1:1 relation where intrinsic and observed size without PSF convolution are equal. 
    PSF effects increase the observed sizes of compact galaxies. This effect is most pronounced for the smallest galaxies at high redshifts ($z \geq 8$). Even at later times ($z \leq 5$), the smaller galaxies in the HD model remain susceptible to such effects.}
    \label{fig:psf-nopsf-r}
\end{figure*}

Here we review systematic offsets between the intrinsic sizes from stellar surface density maps and the half-light sizes from our mock JWST observations. 

The top row of Figure~\ref{fig:psf-nopsf-r} compares the intrinsic size ($\Rstellar$) and the half-light size ($\Rhl$), defined as the observed size without PSF convolution. We show $1\sigma$ and $2\sigma$ kernel density estimate (KDE) contours across our full redshift range: $z = 12$ (red), $z = 8$ (orange), $z = 5$ (green), and $z = 3$ (blue). From left to right, the panels show results for HD, HD-Boost, and \RTnsCRiMHD, respectively. The black dashed line marks the 1:1 relation where intrinsic and half-light size are equal. 

The distributions generally cluster near the 1:1 relation across all models, indicating that half-light sizes broadly trace the underlying stellar density distribution. However, systematic biases emerge at $z \geq 8$ with the strength and redshift evolution of these offsets depending on the feedback model. HD is slightly skewed toward smaller half-light radii relative to stellar radii. Conversely, HD-Boost shows the opposite trend, where half-light radii exceed stellar radii. The \RTnsCRiMHD model follows the 1:1 relation most closely across redshifts. 

In addition to physical mechanisms driving these model-dependent biases, wavelength effects may introduce additional offsets associated with F444W. Shorter wavelengths such as F150W would capture younger and brighter stars, potentially yielding smaller half-light radii. % These findings underscore that inferred galaxy sizes or morphologies, either across different feedback models or redshifts, can be driven as much by observation bias as by genuine physical differences. 

In addition, we show the observed and half-light size bias introduced by PSF convolution. The bottom row of Figure~\ref{fig:psf-nopsf-r} compares the observed galaxy sizes before ($\Rhl$) and after ($\Robs$) applying JWST/NIRCam F444W PSF convolution, following the same panel layout and color scheme as the top row of the figure. Colored vertical dashed lines indicate the physical size of the PSF at each redshift, which imposes an effective resolution floor. For JWST F444W, the physical size of the PSF FWHM is 0.524 kpc at $z=12$, 0.690 kpc at $z = 8$, 0.900 kpc at $z = 5$, and 1.105 kpc at $z = 3$. The PSF acts as a Gaussian blur that defines a minimum resolvable scale, such that any structure or galaxy smaller than $\sim 1.18\,\sigma$, where $\sigma$ is the Gaussian width of the PSF, cannot be resolved. PSF convolution spreads light over this characteristic scale, causing compact sources to appear larger and smoother than their true intrinsic appearance. Therefore, galaxies smaller than this resolution limit produce PSF-dominated profiles, from which size reconstruction is difficult. 

At higher redshifts ($z \geq 8$), galaxy sizes in all three models are subject to instrumental effects due to their smaller sizes. HD and HD-Boost are most affected, with their smallest galaxies at $z = 12$ showing a size increase of up to an order of magnitude. The tightening of contours around the 1:1 line at larger radii ($\Rhl \geq 2~\text{kpc}$) and toward lower redshifts ($z \leq 5$) shows that PSF effects diminish as galaxies become better resolved.

At lower redshifts ($z \leq 5$), HD continues to be affected by PSF effects, as it produces particularly compact galaxies. In contrast, the larger galaxy sizes of the more feedback intense models, HD-Boost and \RTnsCRiMHD, become well-resolved. This reflects the importance of direct comparison between models and observations through forward-modelling methods. This includes instrumental effects such as PSF convolution, which introduce redshift- and model-dependent biases, particularly important for high-redshift galaxies.  

\clearpage
\bibliography{references}
\bibliographystyle{aasjournalv7}
\end{document}